\documentclass[12pt]{article}
\usepackage[english]{babel}
\usepackage{amssymb,amsmath,xcolor,graphicx,xspace,colortbl,rotating} %
\usepackage[title]{appendix}
\usepackage{amsfonts}  
\usepackage{amsmath}
\usepackage{amsthm}
\usepackage{amssymb}  
\usepackage{geometry}  
\usepackage{tabulary}  
\usepackage{graphicx}
\usepackage{caption}     
\usepackage{subcaption}  
\usepackage{hyperref}    
\usepackage{multirow}
\usepackage[flushleft]{threeparttable}
\usepackage[hang,multiple]{footmisc}
\usepackage{chngcntr}
\usepackage{booktabs} 
\usepackage{comment}
\usepackage{csvsimple}
\usepackage{pdflscape}
\usepackage{rotating}
\usepackage{soul}
\usepackage{tikz}
\usepackage{standalone}
\usepackage{pgfplots}
\pgfplotsset{compat=1.17} 
\usepgfplotslibrary{groupplots} 
\usepackage{filecontents}
\usetikzlibrary{calc}
\usepackage[doublespacing]{setspace}
\small\normalsize
\usepackage[round,authoryear]{natbib}
\usepackage{tgpagella} 
\usepackage{subcaption}
\usepackage{hyperref} 
\hypersetup{
    colorlinks=true,
    linkcolor=blue,
    filecolor=magenta,      
    urlcolor=cyan,
    citecolor=blue,
}
\DeclareGraphicsExtensions{.pdf,.eps,.ps,.png,.jpg,.jpeg}
\newtheorem{result}{Result}

\newtheorem{lemma}{Lemma}

\newtheorem{proposition}{Proposition}

\newtheorem{hypothesis}{Hypothesis}
\newtheorem*{subhypothesis*}{Hypothesis 1(a)}

\renewcommand{\thefootnote}{\fnsymbol{footnote}}

\newcommand\smallO{
  \mathchoice
    {{\scriptstyle\mathcal{O}}}
    {{\scriptstyle\mathcal{O}}}
    {{\scriptscriptstyle\mathcal{O}}}
    {\scalebox{.7}{$\scriptscriptstyle\mathcal{O}$}}
  }

\def \baselinestretch{1.5}
\title{Participation, selection and indicative bidding in auctions with costly entry\thanks{\protect\linespread{1}\protect\selectfont We are grateful to Martin Dufwenberg, Lana Friesen, Jacob Goeree, Simon Loertscher as well as audiences at Macquarie University, Monash University, the University of Adelaide, the AGORA Workshop on Behavioral Economics, the Conference on Contests: Theory and Evidence, ANZWEE, APESA, APIOC, and ASSA for helpful comments and suggestions. All errors remain ours. }}

\author{
Changxia Ke\thanks{\protect\linespread{1}\protect\selectfont Changxia Ke: School of Economics and Finance, Queensland University of Technology, QLD 4000, Australia. E-mail: changxia.ke@qut.edu.au}
\and Greg Kubitz\thanks{\protect\linespread{1}\protect\selectfont Greg Kubitz: School of Economics and Finance, Queensland University of Technology, QLD 4000, Australia. E-mail:kubitz@qut.edu.au}
\and Yang Liu\thanks{\protect\linespread{1}\protect\selectfont Yang Liu: School of Economics, Finance and Marketing, RMIT University, VIC 3000, Australia. E-mail: laura.liu@rmit.edu.au}
}
\date{\today}

\begin{document}
\maketitle
\begin{center}
\end{center}
\begin{abstract} 
\thispagestyle {empty}
\onehalfspacing
\def \baselinestretch {1.1}\footnotesize


We study auctions with costly entry in which bidders have incomplete information about their affiliated private valuations prior to entry. We examine whether indicative bidding--a mechanism requiring non-binding preliminary bids before entry--can stimulate participation and improve entrant selection, and we compare its performance with unrestricted and capped entry in a controlled laboratory experiment. When entry costs are high, indicative bidding generates significantly more revenue, primarily by increasing participation beyond theoretical predictions. When entry costs are low, its predicted revenue advantage is attenuated by higher-than-predicted selection inefficiency.

\vspace{10pt}

\vspace{10pt}


\end{abstract}

\newpage 

\setcounter{page}{1}
\pagenumbering{arabic}
\renewcommand{\thefootnote}{\arabic{footnote}}

\section{Introduction}
Developing bids for complex assets often requires bidders to undertake costly due diligence to assess value. Such entry costs arise in settings ranging from company takeovers--where buyers hire professional teams to evaluate targets--to government procurement and natural resource auctions, where bidders invest in market research or site inspections. These costs are ultimately borne by sellers through reduced participation and, consequently, lower expected revenues.\footnote{An example local to the researchers unfolded in 2020 when Deloitte oversaw a takeover auction for Virgin Australia. According to \cite{khadem2020shortlist}, out of 20 initial potential bidders, at least 8 indicative bids were lodged, but only 4 were shortlisted. The importance of the size of the shortlist was highlighted by `Brookfield who reportedly pulled out because it did not want to be part of a group of more than two bidders within a short time frame.'} While limiting the number of bidders can mitigate the impact of these costs by reducing anticipated competition and encouraging entry, doing so requires a selection mechanism when potential demand exceeds the entry cap. The effectiveness of this selection mechanism is crucial: selecting bidders with higher expected values can improve both seller revenue and allocative efficiency. This paper provides novel experimental evidence on how indicative bidding performs as an entry-selection mechanism in auctions with costly entry. Importantly, it is the first to disentangle the mechanism's effectiveness into two channels: increased participation and improved selection efficiency. 

Indicative bidding--widely used in practice--shortlists bidders based on non-binding expressions of interest, typically in the form of indicative bids or bid ranges.\footnote{See examples in \cite{hansen2001auctions}, \cite{boone2007firms} and \cite{gentry2019entry}, as well as discussions in \cite{kagel2008indicative} and \cite{quint2018theory}.}  Although these preliminary bids do not constrain subsequent bidding, they create strategic incentives that can facilitate efficient selection. Bidders face a trade-off: conditional on entry, a higher indicative bid increases the probability of being shortlisted for the auction, while a lower indicative bid decreases the expected competition. High-value bidders therefore have stronger incentives to seek entry, whereas low-value bidders prefer to enter only when competition is limited. Despite its prevalence, there is little causal evidence that indicative bidding simultaneously encourages participation and mitigates selection inefficiencies.

We study this question by comparing indicative bidding (\textit{Ind}) with two alternative entry mechanisms: unrestricted entry (\textit{Unr}), in which all interested bidders may enter, and restricted entry (\textit{Res}) with random selection, in which entry is capped and excess demand is resolved randomly. The indicative bidding mechanism combines an entry cap with endogenous selection, with higher expressions of interest receiving priority in shortlisting.

To guide the experimental analysis, we develop a theoretical framework based on \cite{quint2018theory}. A set of risk-neutral bidders compete for a single indivisible asset. Each bidder has a private value that is unknown initially; instead, they observe a private signal and learn their full valuation only after entering the auction and paying a fixed entry cost that is identical across entrants. A bidder's private value is the sum of their initial private signal that is independently drawn from a uniform distribution and a common value component that is revealed after entry.\footnote{The common value assumption represents a special case of highly correlated information acquired after entry. We adopt this because it both mirrors the information structure of our leading examples \citep{li2009entry, aradillas2013identification} and is a setting in which efficient selection is particularly critical for auction revenue.} The auction proceeds as a second-price sealed-bid auction with no reservation price.\footnote{While the first-price auction is commonly used in these settings, the second-price auction guarantees the existence and uniqueness of a symmetric equilibrium for our theoretical framework of the experimental settings. Both auction formats provide similar quantitative predictions for revenue and surplus for parameters used in our experiment, see \cite{quint2018theory}, Tables 1 and 3.}

To quantify revenue loss due to costly entry, we compare expected revenue under each entry mechanism with the expected revenue of the zero-entry-cost benchmark. To identify the channels underlying this revenue loss, we decompose it into two components: a \textit{participation effect} and a \textit{selection effect}.  The \textit{participation effect} captures revenue loss due to potential bidders choosing not to participate in the auction due to costly entry, while the \textit{selection effect} captures the loss arising from the mechanism failing to admit the highest-value interested bidders into the auction stage. 
 
Our theoretical predictions first characterize the revenue trade-off induced by restricting entry. Reducing the maximum number of bidders increases expected participation by lowering anticipated competition, thereby mitigating the \textit{participation effect}. At the same time, it reduces the likelihood that the highest-value interested bidders are selected, increasing the \textit{selection effect}. Which force dominates depends on the level of entry costs. When entry costs are sufficiently high, the reduction in \textit{participation effect} outweighs the increase in the selection effect, so restricting entry raises revenue and \textit{Res} outperforms \textit{Unr}. When entry costs are sufficiently low, the opposite holds and restricting entry reduces revenue. We then show that allowing bidders to indicate two levels of interest to enter always reduces the \textit{participation effect}. Moreover, for the value distributions and the number of bidders considered in the experimental setup, indicative bidding also reduces the \textit{selection effect} relative to random selection. As a result, \textit{Ind} is predicted to generate at least as much revenue as \textit{Res} for any entry cost and more than \textit{Unr} for sufficiently high entry costs.

We test these predictions in a $2\times 3$ laboratory experiment that varies both the entry mechanism (\textit{Unr}, \textit{Res}, and \textit{Ind}) and the level of entry cost (\textit{low-cost} and \textit{high-cost} conditions). Consistent with theoretical predictions, \textit{Ind} and \textit{Unr} generate higher average revenue than \textit{Res} when entry costs are low. When entry costs are high, \textit{Ind} generates higher average revenue than both \textit{Unr} and \textit{Res}, even though \textit{Res} is predicted to yield the same average revenue as \textit{Ind}. Overall, \textit{Ind} delivers weakly higher average revenue than the two alternative mechanisms across entry-cost conditions. 

Despite this overall alignment with the theory, we also observe three deviations from equilibrium. First, non-equilibrium entry behavior leads to a higher \textit{selection effect} under both \textit{Res} and \textit{Ind}, which helps explain why \textit{Res} and \textit{Ind} do not outperform \textit{Unr} in the \textit{low-cost} condition. Second, participation in \textit{Ind} is higher than predicted, further reducing the \textit{participation effect} and contributing to \textit{Ind}'s superior performance relative to \textit{Res} under high entry costs. Third, bidders overbid on average in the second-price auction across treatments, a pattern consistent with existing experimental evidence that does not materially affect cross-treatment comparisons.

We also examine bidders' profits and social welfare--defined as the sum of auction revenue and bidders' profits--across mechanisms. \textit{Ind} generates higher social welfare than \textit{Unr} under both cost conditions, while \textit{Res} improves welfare relative to \textit{Unr} when entry costs are high. Although \textit{Ind} raises revenue relative to \textit{Res}, the welfare gains are limited, as the additional revenue largely comes at the expense of bidders' profits and is driven primarily by increased participation rather than improved selection.

 Our paper contributes to the literature on auctions with costly and endogenous entry, which emphasizes the role of participation decisions in shaping auction outcomes. When bidders must incur costly due diligence, unrestricted competition can reduce revenue and welfare by discouraging entry. As \citet[p.~209]{milgrom2004putting} describes ``Buyers are naturally reluctant to begin an expensive and time-consuming evaluation of an asset when they believe they are unlikely to win at a favorable price." Entry costs therefore generate coordination problems that are absent in standard auction models with exogenous participation. When bidders are ex-ante symmetric and learn their values only after entering, coordination failures reduce participation and expected revenue as the number of potential bidders increases \citep{levin1994equilibrium}. When bidders observe informative signals before entry, entry cutoff points increase with the number of potential bidders, which can similarly reduce revenue and social welfare \citep{samuelson1985competitive, li2009entry}. These papers establish that costly entry weakens the link between competition and revenue, focusing on environments without explicit entry-selection mechanisms. 

One response to inefficient entry is the use of entry fees or reserve prices to screen out low-value or high-cost bidders.\footnote{An entry fee is an imposed payment that transfers surplus to the seller, whereas an entry cost is a real participation cost borne by bidders.} Such instruments can increase seller revenue and welfare in theory \citep{menezes2000auctions, lu2009auction,lu2010entry}, but may require upfront, non-refundable payments that may be impractical in many applied settings. Another response is to use two-stage auction mechanisms (i.e., those that include a pre-qualification stage) that explicitly coordinate participation. A simple form is a fixed cap on the number of bidders. 
When bidders have no information before entry, \cite{moreno2011auctions} show that an entry cap can improve social welfare and, when combined with an entry fee, increases seller revenue. Using the \cite{quint2018theory} framework with a single positive entry message, we extend this logic to environments in which bidders observe private signals prior to entry, showing theoretically that an entry cap can increase revenue when entry costs are sufficiently high even without entry fees. Our experimental results, however, reveal that these revenue gains fail to materialize in practice due to selection inefficiencies arising from non-equilibrium entry behavior, an effect that cannot be identified in purely theoretical analysis.

Indicative bidding introduces an additional margin of selection by allowing bidders to signal varying levels of interest before entry. Although a fully separating equilibrium generally fails to exist with non-binding signals (see \cite{ye2007indicative}), \cite{quint2018theory} demonstrate that there is a symmetric semi-separating equilibrium in which indicative bidding partially sorts bidders by value and improves selection relative to random assignment. Indicative bidding also contrasts with entry rights auctions, in which bidders compete in a first-stage auction for the right to enter the main auction. 
\cite{ye2007indicative} shows that entry rights auctions can induce efficient entry in a symmetric equilibrium. Moreover, \cite{lu2021orchestrating} identify that the revenue maximizing two-stage auction involves message-dependent shortlisting and payment rules so that the number of entrants and their first-stage payments are endogenously determined. 
However, these mechanisms require bidders to make binding, non-refundable payments before acquiring full information about the target, limiting their appeal in settings with costly due diligence. In addition, \cite{kagel2008indicative} provide experimental evidence comparing entry rights auctions and indicative bidding, finding that indicative bidding performs at least as well in selecting high-value bidders while generating higher revenue and fewer bankruptcies. Because no equilibrium predictions for indicative bidding were available within their theoretical framework, selection efficiency could not be assessed relative to theory.\footnote{Relatedly, the “qualifying auction” with an inside bidder studied by \cite{boone2009optimal}--in which a non-binding first stage is used to exclude the bidder with the lowest bid--was experimentally tested by \cite{boone2009risky}. They observe arbitrarily high bidding (a babbling equilibrium) in the non-binding stage, along with lower auction revenue compared to a standard English auction. Crucially, their model assumes costless entry, so unlike our setting, bidding higher in the indicative stage does not increase expected costs. Other experimental work on auctions with endogenous entry has focused on testing revenue equivalence or winner's curse under free or exogenous entry \citep{ivanova2008revenue,cox2001endogenous,palfrey2008endogenous}.}

Finally, our paper also relates to sequential mechanisms, such as those studied by \cite{bulow2009sellers} and \cite{roberts2013should}, which allow sellers to solicit bids sequentially from potential bidders. Each potential bidder's entry decision depends on the current bid, and therefore bidders may want to place a jump bid to discourage future entry. Although aggressive early bids have a direct effect of increasing auction revenue, by deterring future entry, expected revenue may actually decrease. 
\cite{lu2021orchestrating} show that the revenue maximizing multi-stage shortlisting rule involves sequentially engaging the potential bidders with the highest pre-entry value and charging them non-refundable entry fees. 

Our experiment directly studies how entry-selection mechanisms shape participation, selection, and welfare in environments with costly information acquisition. Our findings help explain why indicative bidding performs well in practice, and clarify the channels--participation versus selection--through which entry-selection rules affect auction outcomes. By designing the indicative bidding stage to mirror the restricted theoretical message space based on \cite{quint2018theory}, we directly compare the observed behavior to equilibrium benchmarks and show that while indicative bidding improves revenue relative to random selection, the gains arise primarily through increased participation (\textit{participation effect}) rather than improved selection efficiency (\textit{selection effect}).


\section{Theoretical Framework}\label{sec:theory}

Our model considers a two-stage auction with costly entry and follows the information structure in  \cite{quint2018theory}. There is a set of $N\ge 3$ risk-neutral potential bidders, each with private value $V_i$ for a single indivisible asset. Potential bidders do not initially know $V_i$, but instead have a private signal $S_i$. Each bidder’s valuation is given by $V_i = S_i+ T$, where $T$ is a common value component learned upon entering the auction and paying an entry cost $c$. We assume that the bidders' signals are independently and uniformly distributed on $[0,\bar S]$, and $\{S_1,\ldots,S_N,T\}$ are mutually independent. We also take $c\in(0,\bar S+\mathbb{E}[T])$ which rules out cases in which bidders' participation decisions do not depend on the auction mechanism used. The equilibrium-relevant parameters of this setting can be summarized by $\mathcal{X}=(\bar S,\mathbb{E}[T],N,c)$, which we call the auction environment. 

We consider two-stage mechanisms consisting of a communication stage and an auction stage. In the communication stage, potential bidders privately observe $S_i$ and send one message from a finite set $m\in\{0,1,\ldots,\overline{M}\}$, indexed by nonnegative integers.\footnote{ Although this set can be countably infinite in principle, in any symmetric equilibrium only a finite number of messages is ever used (see \cite{quint2018theory}, Lemma 1).} Sending $m=0$ corresponds to opting out, while sending positive messages indicates willingness to participate. Bidders are admitted to the auction stage through a precommitted selection process that caps the number of bidders at $n\ge 2$. Those who send the highest positive messages are selected sequentially until all $n$ positions are filled, with ties broken randomly. The auction stage is a standard second-price auction with no reservation price. A two-stage mechanism is therefore characterized by the entry cap $n$ and the number of positive messages available $\overline{M}$.

For any environment $\mathcal{X}$, there exists an essentially unique symmetric equilibrium for each two-stage mechanism $\mathcal{M}=(n,\overline M)$.\footnote{The formal details of the assumptions required for the existence of a symmetric equilibrium are given in Assumptions 1 and 2 in \cite{quint2018theory}. In our setting, the existence and essential uniqueness follow from three conditions: the  mutual independence of $\{S_1,\ldots,S_N,T\}$, the uniform distribution of each $S_i$, and the fact that $T$ is a common value.} In this equilibrium, the communication stage is partially separating: potential bidders with higher signals send weakly higher messages. Equilibrium strategies can be described by cutoff signals $\alpha_m$, where bidders with $S_i\in[\alpha_m,\alpha_{m+1}]$ optimally choose message $m$, and are indifferent between messages $m$ and $m+1$ at cutoff $\alpha_m$.

\subsection{Revenue loss, \textit{participation} and \textit{selection} effects}\label{subsec:Def_PE_SE}

For a given environment $\mathcal{X}$, we evaluate the performance of a mechanism $\mathcal{M}$ in the presence of entry costs by comparing its expected revenue, $\Pi(\mathcal{M}, \mathcal{X})$, to the expected revenue in the same environment without entry costs. When entry is costless ($c=0$), all bidders enter, restricting entry is unnecessary, and auction revenue equals the second-highest valuation among all potential bidders. Let $\Pi(\mathcal{X}_0)$ denote the expected revenue in the zero-entry-cost environment $\mathcal{X}_0$, which is identical to $\mathcal{X}$ except that $c=0$. We then define the total revenue loss from costly entry under the mechanism $\mathcal{M}$ as $RL(\mathcal{M},\mathcal{X}) \equiv \Pi(\mathcal{X}_0) - \Pi(\mathcal{M},\mathcal{X})$.

To isolate the \textit{participation} and \textit{selection} effects, we consider a counterfactual scenario in which all potential bidders who send a positive entry message are allowed to enter the auction stage, irrespective of the mechanism's entry cap or selection rule. Let $\hat\Pi(\alpha_0(\mathcal{M}),\mathcal{X})$ denote the expected revenue in this counterfactual scenario.\footnote{We use the notation $\alpha_0(\mathcal{M})$ to represent the equilibrium cutoff signal of bidders that are indifferent between opting out with $m=0$ or participating with $m=1$ given mechanism $\mathcal{M}$. This cutoff signal is sufficient to characterize this hypothetical revenue given $\mathcal{X}$.} Then, for a given mechanism $\mathcal{M}$ and environment $\mathcal{X}$,  we define the \textit{participation effect} and the \textit{selection effect}, respectively, as $PE(\mathcal{M},\mathcal{X})\equiv  \Pi(\mathcal{X}_0) - \hat\Pi(\alpha_0(\mathcal{M}),\mathcal{X})$ and $SE(\mathcal{M},\mathcal{X}) \equiv \hat\Pi(\alpha_0(\mathcal{M}),\mathcal{X}) - \Pi(\mathcal{M},\mathcal{X})$. The \textit{participation effect} quantifies the expected revenue loss due to potential bidders who optimally choose to opt out due to costly entry, while the \textit{selection effect} captures the expected revenue loss arising from the mechanism's failure to admit the highest-value participating bidders into the auction stage. By construction, the total revenue loss is additively decomposed into these two components.

In the remainder of this section, we first analyze how restricting entry affects revenue when bidders can send only two messages in the communication stage (“opt in” or “opt out”). We then study how allowing for an additional positive entry message alters participation, selection, and overall revenue performance. The proofs for all the results in this section are in \autoref{appendix:proofs}.

\subsubsection{Entry restriction} \label{subsec:Equ}

Consider a two-stage mechanism where potential bidders can only opt in or opt out of the auction ($\overline M = 1$). The symmetric equilibrium is then characterized by a single cutoff signal ($\alpha_0$) in the communication stage. For a given environment and an entry cap $n$, this cutoff signal satisfies: 
\begin{equation} 
 p_0 (\alpha_0 + \mathbb{E}[T]-c) - c \sum_{k=1}^{N-1} p_k \min\left\{1,\frac{n}{k+1}\right\} = 0, \label{RES cut-off}
\end{equation}
where $p_k = {N-1 \choose k} \left(\frac{\bar S-\alpha_0}{\bar S}\right)^k \left(\frac{\alpha_0}{\bar S}\right)^{N-1-k}$ is the probability that $k$ other potential bidders will choose $m=1$. Equation \ref{RES cut-off} implies that a potential bidder with the cutoff signal $\alpha_0$ is indifferent between opting in and opting out of the auction. This bidder would win the auction only if no other potential bidders opt in, which happens with probability $p_0$. The bidder's expected profit in this case is $\alpha_0 + \mathbb{E}[T] - c$. When $k\ge 1$ other potential bidders opt in, the potential bidder with the cutoff signal is selected with probability $\min\left\{1,\frac{n}{k+1}\right\}$, pays the entry cost, but does not win the auction. A potential bidder who opts in faces a lower probability of being selected into the auction stage whenever more than $n$ potential bidders choose to participate. 

Lemma~\ref{lemma:entry restriction cutoff} shows that $\alpha_0$ decreases as $n$ decreases, implying that a more restrictive entry induces participation of bidders with lower initial signals. 

\begin{lemma} \label{lemma:entry restriction cutoff}
For a given environment and the set of mechanisms with $\overline M = 1$, a mechanism with a lower entry cap $n$ has a lower cutoff signal $\alpha_0$.
\end{lemma}

Although an entry restriction encourages more potential bidders to opt in, it also increases the likelihood that some interested bidders will be excluded from the auction stage when demand exceeds the entry cap. As a result, decreases in revenue loss due to the \textit{participation effect} may be offset by increases in the \textit{selection effect}. Proposition \ref{proposition:entry restriction} formalizes this trade-off. 

\begin{proposition} \label{proposition:entry restriction}
For a given environment and a set of mechanisms with $\overline M = 1$, a mechanism with a lower maximum number of bidders ($n$) decreases the \textit{participation effect} and increases the \textit{selection effect}.
\end{proposition}

The revenue consequences of restricting entry depend on the relative strength of the \textit{participation} and \textit{selection} effects. When the entry cost is low, most potential bidders are willing to opt in even with unrestricted entry. In this case, restricting entry yields only a small reduction in the \textit{participation effect} while substantially increasing the \textit{selection effect}, leading to higher expected revenue loss. By contrast, when the entry cost is high, participation is limited under unrestricted entry. Restricting entry then generates a large reduction in the \textit{participation effect} while inducing only a modest increase in the \textit{selection effect}, resulting in a lower expected revenue loss. 

Proposition~\ref{proposition:PE vs SE} shows that when the cost of entry is high enough, it is revenue maximizing to restrict the number of bidders to $n=2$. Whereas when the entry cost is sufficiently low, placing no restriction on the number of bidders maximizes revenue.

\begin{proposition} \label{proposition:PE vs SE}
Fix an environment other than the entry cost and consider the set of mechanisms with $\overline M = 1$.
\begin{itemize}
    \item[(a)] For sufficiently low entry costs, any mechanism that restricts the number of bidders ($n<N$) yields a higher expected revenue loss than the unrestricted mechanism with $n=N$. \label{prop:1a}
    \item[(b)] For sufficiently high entry costs, the mechanism that restricts the number of bidders to $n=2$ yields a lower expected revenue loss than the mechanisms that allow $n>2$ bidders.\label{prop:1b}
\end{itemize}
\end{proposition}

\subsubsection{Indicative bidding}

We now consider the impact of adding an additional entry message ($\overline M = 2$) to the communication stage of the mechanism. With two entry messages (and three total messages), there are up to two cutoff signals $\alpha_0$ and $\alpha_1$ in the symmetric equilibrium. The cutoff signal for potential bidders who are indifferent between opting in with the entry message $m=1$ and opting out is given by the following equation:
\begin{equation}  \label{IND cut-off 1}
p_{0,0}(\alpha_0 + \mathbb{E}[T]-c)
- c \sum_{j=0}^{N-1} \sum_{k=1}^{N-1-j} p_{k,j} \min\left\{1,\frac{n-j}{k+1}\right\} = 0,
\end{equation}
where $p_{k,j} = {N-1 \choose j} {N-1-j \choose k} \left(\frac{\bar S - \alpha_1}{\bar S} \right)^j \left(\frac{\alpha_1 -\alpha_0}{\bar S} \right)^k \left(\frac{\alpha_0}{\bar S} \right)^{N-1-j-k}$ is the probability that $k$ other potential bidders choose $m=1$ and $j$ others choose $m=2$.  

Potential bidders will prefer to send a higher entry message $m=2$ when they want to increase their chances of entering the auction despite knowing that they face competition. The cutoff signal for which a bidder is indifferent between choosing $m=1$ and $m=2$ must satisfy 
\begin{align} \label{IND cut-off 2}
\sum_{k=n}^{N-1} p_{k,0} \left(1 - \frac{n}{k+1} \right) & \left( \frac{\alpha_1-\alpha_0}{n} - c\right) \nonumber\\
& - c \sum_{j=1}^{N-1} \sum_{k=\max\{0,n-j\}}^{N-1-j} p_{k,j} \left(\min\left\{1,\frac{n}{j+1}\right\} - \frac{\max\{0,n-j\}}{k+1} \right) = 0.
\end{align}
The first term is the expected benefit of increasing the probability of entering when no other potential bidders have a signal above $\alpha_1$, and the second term is the expected cost of increasing the chance of entering the auction when at least one potential bidder has a signal above $\alpha_1$. In an equilibrium where $m=2$ is used with positive probability, the cutoff signals must jointly satisfy (\ref{IND cut-off 1}) and (\ref{IND cut-off 2}). When the entry cost is high enough, these equations do not have a solution for $\alpha_1, \alpha_0\in[0,\bar S]$; hence, in equilibrium, potential bidders choose only $m=0$ and $m=1$.\footnote{In this case, the first term of (\ref{IND cut-off 2}) will be less than the second term for all $\alpha_1,\alpha_0$ that satisfy (\ref{IND cut-off 1}). For higher entry costs, a potential bidder needs a larger expected gain to prefer $m=1$ over $m=0$. This raises the cutoff $\alpha_0$ satisfying (\ref{IND cut-off 1}) for a given $\alpha_1$. The higher $\alpha_0$ reduces the expected benefit from choosing $m=2$ instead of $m=1$ for a given $\alpha_1$, and higher entry costs increase the expected cost of entering more often. To keep (\ref{IND cut-off 2}) satisfied, $\alpha_1$ must also rise. For sufficiently high entry cost, (\ref{IND cut-off 1}) and (\ref{IND cut-off 2}) can only be satisfied by $\alpha_1 > \bar S$ and therefore, in equilibrium, $m=2$ is not chosen.}

When there is a second entry message, a potential bidder sending $m=1$ is less likely to be selected to enter the auction when facing competition. This increases the expected value of choosing $m=1$ relative to $m=0$ and reduces $\alpha_0$. Since, for a fixed environment, the \textit{participation effect} is only affected by the change in the cutoff signal $\alpha_0$, the following proposition shows that the \textit{participation effect} will be strictly lower when the second entry message is used in equilibrium.

\begin{proposition}\label{proposition:indicative bidding}
    For any environment and entry restriction, the cutoff value $\alpha_0$ and the \textit{participation effect} are weakly lower when $\overline M = 2$ than when $\overline M = 1$. These reductions are strict when $\alpha_1<\bar S$.
\end{proposition}

\begin{figure}[htbp]
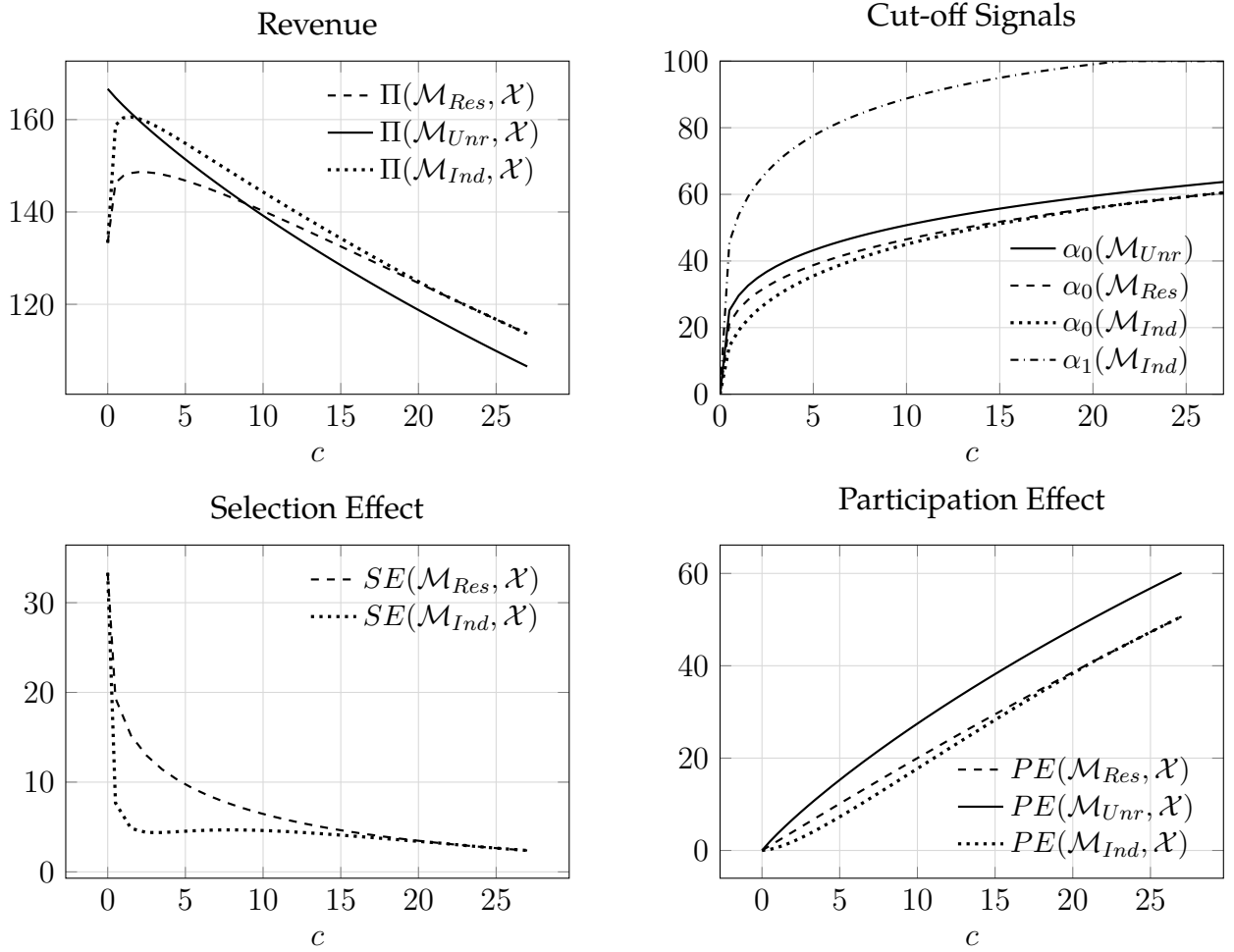

    \centering
\includestandalone[width=\textwidth]{Combined_Graph}    \caption{Comparison of revenue, equilibrium cutoff signals, \textit{selection effect} and \textit{participation effect} for the three two-stage mechanisms $\mathcal{M}_{Unr} = (5,1)$, $\mathcal{M}_{Res} = (2,1)$ and $\mathcal{M}_{Ind} = (2,2)$ under the environment $\mathcal{X} = (100,100,5,c)$ and plotted against different values of $c$, which include the \textit{low-cost} $c=5$ and \textit{high-cost} $c=25$ experimental conditions.}
    \label{fig:cutoffs and revenue}
\end{figure}

Although the \textit{participation effect} is always reduced when a second message is added, this is not necessarily true for the \textit{selection effect}.\footnote{Furthermore, when the entry cost is sufficiently small, a large \textit{selection effect} leads to more expected revenue loss than the unrestricted mechanism even when $\overline M=2$. Therefore, similar to the result of Proposition~\ref{proposition:entry restriction}, an unrestricted entry mechanism with $n=N$ will lead to less revenue loss than a restricted entry mechanism with multiple entry messages.} The availability of a second entry message will increase the chance that a high-value bidder with $S_i\in(\alpha_1,\bar S)$ is selected to enter the auction when $\alpha_1< \bar S$. However, the second entry message also lowers $\alpha_0$, increasing the chance that a low-valued bidder chooses $m=1$ and is selected to enter over a higher valued bidder who also chose $m=1$. For the parameters used in our experimental setting, the \textit{selection effect} does decrease when a second entry message is added. As shown in Figure~\ref{fig:cutoffs and revenue}, the second entry message reduces both the \textit{participation effect} and the \textit{selection effect} and therefore leads to less revenue loss for all entry costs given the other parameters used in the experiment.

\section{Experimental design and  hypotheses}\label{sec:design}

Based on the theoretical predictions given in Section \ref{sec:theory}, our $2\times3$ experimental design aims to compare the performance of unrestricted entry $(\mathcal{M}_{Unr})$, restricted entry $(\mathcal{M}_{Res})$, and indicative bidding $(\mathcal{M}_{Ind})$ mechanisms in both \textit{low-cost} ($c=5$) and \textit{high-cost} ($c=25$) conditions. For each treatment, we fix the environment (other than the entry cost) so that $N=5$, $\bar S = 100$ and $\mathbb{E}[T]=100$.\footnote{As described in Section~\ref{sec:theory}, the two-stage entry mechanisms can be summarized by the number of bidders allowed to enter, $n$, and the number of positive messages available, $\overline M$: $\mathcal{M}=(n,\overline M)$. The mechanisms tested are $\mathcal{M}_{Unr} = (5,1), \mathcal{M}_{Res} = (2,1)$, and $\mathcal{M}_{Ind} = (2,2)$.} 

In the experiment, each two-stage game unfolds as follows. The computer first randomly draws an integer from a uniform distribution on [0,100] for each participant independently as their initial signal $S_i$. Having observed their private signal, the participants choose an entry message. Given the chosen messages, the entrants are selected according to the treatment-specific mechanism. The selected participants are advanced to the auction stage, pay the entry cost $c$ and learn the additional common value $T$ which follows a discrete distribution with probability 1/2 taking the value of 100 and probability 1/4 taking the value of either 0 or 200 respectively.\footnote{The common value $T$ is determined for all participants (within the same experimental group) by randomly assigning one of two numbers--100 and 0--to two cards independently. The sum of the two numbers assigned to the two cards determines the common value. Hence, the common value has 25\% being 0, 50\% being 100, and 25\% being 200. See Appendix~\ref{appendix:instructions} for details on how this was implemented in the experiment.} After learning the common value component, a second-price sealed-bid auction is conducted. The entrants must place their bid without knowing the number of other entrants against whom they are bidding. When there is only one entrant, that entrant always wins the auction and pays zero. Once the auction is completed, all participants receive complete feedback on each participant's initial signal, entry decision, bid (if selected to enter), and payoff in that round.\footnote{A history table that includes the number of participants who chose to enter in each round, as well as the winner's valuation, bid, and profit in all previous rounds, is also displayed on every decision screen after round 1.}

We employ a hybrid between-subject and within-subject experimental design. In each session, participants complete 20 rounds of the two-stage game under the same mechanism, with no carryover value between rounds. The entry cost is set to $5$ (\textit{low-cost} condition) in rounds 1-10 and to $25$ (\textit{high-cost} condition) in rounds 11-20. Participants are randomly assigned to groups of five in round 1 and remain in the same group for rounds 1-10. In round 11, they are randomly rematched (coinciding with the change in the entry cost) and remain in their new group for rounds 10-20. At the end of the session, one round is randomly selected for payment.\footnote{To cover the entry cost, participants were given 30 experimental currency (i.e., EC) each round as their initial endowment. Hence, the final payment also included the (remaining) endowment in the selected payment round. Their earnings from the payment round were converted to RMB using an exchange rate of 1 EC = 1.5 RMB.} 

The experiment was programmed using oTree \citep{chen2016otree} and conducted at Zhejiang Gongshang University in China. We recruited 360 students across 12 sessions, with four sessions per treatment and 30 participants in each session.\footnote{To help the participants understand the experimental procedure, at the beginning of each session, a video that summarizes the experimental instructions was played after they read the printed instructions. Participants were also given control questions which they had to answer fully correctly before entering the main part of the experiment. Please see Appendix~\ref{appendix:instructions} for the experimental instructions and relevant screenshots.} The sessions lasted approximately 1.5 hours. Average earnings were around 75 RMB, inclusive of a 30 RMB show-up fee.

\subsection{Predictions and Hypotheses}
\label{subsec:hypotheses}
Table \ref{tab: prediction} presents the predicted cutoff for each entry message, the total revenue losses due to costly entry, along with the predicted \textit{selection} and \textit{participation} effects, auction revenue, bidders' profit, and social welfare under each mechanism.

\vspace{5mm}
\begin{table}[ht!]
\begin{center}\footnotesize
\begin{threeparttable}
  \caption{Theoretical Predictions}
  \label{tab: prediction}
    \begin{tabular}{lcccrccc}
\toprule
&\multicolumn{3}{c}{\textbf{Low-cost (c=5)}}&&\multicolumn{3}{c}{\textbf{High-cost(c=25)}} \\
& \textbf{Unr} & \textbf{Res} & \textbf{Ind} & & \textbf{Unr} &\textbf{Res} & \textbf{Ind} \\
\cmidrule{2-4}\cmidrule{6-8} 
    $\alpha_0$ & 43.23 & 38.74 & 35.50 &       & 62.62 & 59.28 & 59.28\\
    $\alpha_1$ &       &       & 77.59 &       &       &       &  -\\
\midrule
    Revenue Loss (RL) & 15.27 & 19.87 & 10.89 &       & 56.77 & 49.93 & 49.93  \\
\ \ \ \textit{participation effect} & 15.27 & 10.12 & 9.77 &       & 56.77 & 47.28 & 47.28 \\
\ \ \ \textit{selection effect} & 0     & 9.75 & 1.12 &       & 0     & 2.65 & 2.65 \\
\midrule
    Social welfare & 167.10 & 167.74 & 172.02 &       & 121.96 & 131.00 & 131.00  \\       
\ \ \ Auction revenue & 151.39 & 146.79 & 154.86 &       & 109.89 & 116.73 &116.73  \\
\ \ \ Bidders' profit & 15.71 & 20.95 & 17.16&       & 12.07 & 14.27 & 14.27  \\
    \bottomrule
    \end{tabular}%
\begin{tablenotes}[flushleft]
\item Note: Theoretical predictions are calculated using the parameters from the experiment. Specifically, the three two-stage mechanisms $\mathcal{M}_{Unr} = (5,1)$, $\mathcal{M}_{Res} = (2,1)$ and $\mathcal{M}_{Ind} = (2,2)$ under the environment $\mathcal{X} = (100,100,5,c)$, with $c=5$ and  $c=25$ in \textit{low-cost} and \textit{high-cost} conditions, respectively. $\alpha_0$ and $\alpha_1$ are the cutoffs for message $m=1$ and $m=2$, respectively. Revenue loss (RL) is the difference in revenue between \textit{Unr}, \textit{Res}, or \textit{Ind} and the auction with zero entry cost. 
\end{tablenotes}
\end{threeparttable}
\end{center}
\end{table}%

As indicated by Proposition~\ref{proposition:PE vs SE} (and shown in Table~\ref{tab: prediction}), \textit{Unr} generates more revenue than \textit{Res} when the entry cost is low; and when the entry cost is high, the reverse order holds. Moreover, from Figure~\ref{fig:cutoffs and revenue}, \textit{Ind} will lead to weakly higher revenues than \textit{Res}  for either cost condition. This leads to the following revenue rankings:

\begin{hypothesis}[\textbf{Revenue ranking}]
When the entry cost is low, $\Pi(\mathcal{M}_{Ind},\mathcal{X}) > \Pi(\mathcal{M}_{Unr},\mathcal{X}) > \Pi(\mathcal{M}_{Res},\mathcal{X})$; when the entry cost is high,  $\Pi(\mathcal{M}_{Ind},\mathcal{X}) = \Pi(\mathcal{M}_{Res},\mathcal{X}) > \Pi(\mathcal{M}_{Unr},\mathcal{X})$.
\end{hypothesis}

Proposition \ref{proposition:entry restriction} and Proposition \ref{proposition:PE vs SE} characterize the direction and magnitude of changes in the \textit{participation} and \textit{selection} effects with respect to the entry restriction, $n$, when there is one positive entry message ($\overline{M}=1$). Proposition \ref{proposition:indicative bidding} further shows that the \textit{participation effect} (weakly) decreases when a second message is added. Figure~\ref{fig:cutoffs and revenue} shows that the \textit{selection effect} also decreases weakly. Comparisons of these effects across the three mechanisms and two cost conditions are summarized by the following hypotheses. 

\begin{hypothesis}[\textbf{Participation effect}]
When the entry cost is low, $PE(\mathcal{M}_{Ind},\mathcal{X}) < PE(\mathcal{M}_{Res},\mathcal{X}) < PE(\mathcal{M}_{Unr},\mathcal{X})$; when the entry cost is high, $PE(\mathcal{M}_{Ind},\mathcal{X}) = PE(\mathcal{M}_{Res},\mathcal{X}) < PE(\mathcal{M}_{Unr},\mathcal{X})$.
\end{hypothesis}

\begin{hypothesis}[\textbf{Selection effect}]
When the entry cost is low, $0 = SE(\mathcal{M}_{Unr},\mathcal{X}) < SE(\mathcal{M}_{Ind},\mathcal{X}) < SE(\mathcal{M}_{Res},\mathcal{X})$; when the entry cost is high, $0 = SE(\mathcal{M}_{Unr},\mathcal{X}) < SE(\mathcal{M}_{Ind},\mathcal{X}) = SE(\mathcal{M}_{Res},\mathcal{X})$.
\end{hypothesis}

By limiting the number of bidders, both the restricted entry and indicative bidding mechanisms increase bidders' expected profit by reducing the number of participants who incur the entry cost.\footnote{See Equations (\ref{equation:profit res}) and (\ref{equation:profit ind}) in Appendix~\ref{appendix:proofs}.} This increase in bidders' surplus contributes to higher social welfare. When entry costs are low, the gain in bidders' profit largely offsets the reduction in seller's revenue from restricting entry, so the unrestricted and restricted mechanisms yield similar levels of social welfare (see Table~\ref{tab: prediction}). When entry costs are high, restricting entry raises both seller's revenue and bidders' profit, resulting in higher social welfare. In addition, when multiple positive messages are used, the indicative bidding mechanism can further increase social welfare by selecting bidders with higher expected valuations into the auction. These considerations lead to the following hypothesis regarding the ranking of social welfare across entry mechanisms.

\begin{hypothesis}[\textbf{Social Welfare}]
When the entry cost is low, $\mathcal{M}_{Ind}$ generates more social welfare than $\mathcal{M}_{Res}$ and $\mathcal{M}_{Unr}$; when the entry cost is high, both $\mathcal{M}_{Ind}$ and $\mathcal{M}_{Res}$ generate more social welfare than $\mathcal{M}_{Unr}$. 
\end{hypothesis}

\section{Experimental Results}\label{sec:results}
In this section, we first compare auction revenue across treatments and examine how differences in revenue are driven by the \textit{participation} and \textit{selection} effects. We then analyze individual entry and bidding behavior to understand the behavioral sources of these effects. Finally, we compare social welfare and bidders' profits across treatments.

\subsection{Auction revenue: \textit{participation} and \textit{selection effects}} \label{subsec:results_rev}
Table \ref{tab: rev} reports multilevel linear regressions of auction revenue under each cost condition, estimated with both session- and group-level random effects. In all specifications, the unrestricted mechanism (\textit{Unr}) serves as the baseline. Model 1 compares the restricted entry (\textit{Res}) and indicative bidding (\textit{Ind}) treatments to \textit{Unr}. Model 2 additionally controls for the second-highest realized value in each group to account for differences in value draws across treatments.\footnote{Participants' valuations are randomly drawn by the computer program in each session of the experiment, and controlling for the second-highest value in each group reduces variations in revenue from these draws.} As the estimates are qualitatively similar across models, we focus on Model 2 in the following discussion. 

In the \textit{low-cost} condition, the revenue in \textit{Res} is 13.91 units lower than in \textit{Unr} ($p< 0.01$) and 10.60 units lower than in \textit{Ind} ($p = 0.04$). The revenue in \textit{Ind} is also slightly lower than in \textit{Unr}, but the difference is not statistically significant ($p= 0.53$). In the \textit{high-cost} condition, the revenue in  \textit{Ind} exceeds that in \textit{Unr} by 25.37 units ($p< 0.01$) and that in \textit{Res} by around 20 units  ($p = 0.03$). Although \textit{Res} is predicted to outperform \textit{Unr} in the \textit{high-cost} condition, the observed difference of 5.65 units is not statistically significant ($p = 0.54$). Overall, these results provide partial support for Hypothesis 1.

\begin{table}[h!]
\centering
\begin{threeparttable}
\footnotesize
    \caption{Multilevel linear regressions on auction revenue} \label{tab: rev}
    \begin{tabular}{lcccc}
    \toprule
          & \multicolumn{2}{c}{\textbf{Model 1}} & \multicolumn{2}{c}{\textbf{Model 2}} \\
\cmidrule{2-5}  & Low-cost & High-cost & Low-cost & High-cost \\
    \midrule
    Unr (Cons.) & 156.3*** & 107.1*** & -1.926 & -30.25*** \\
          & (5.726) & (7.455) & (5.927) & (9.195) \\
    Res   & -15.51* & -2.121 & -13.91*** & 5.653 \\
          & (8.098) & (10.54) & (5.205) & (9.272) \\
    Ind   & -7.217 & 24.79** & -3.313 & 25.37*** \\
          & (8.098) & (10.54) & (5.206) & (9.265) \\
    V$_2$ &       &       & 0.939*** & 0.840*** \\
          &       &       & (0.0276) & (0.0394) \\
    \# of Obs. & 720   & 720   & 720   & 720 \\
    \# of Groups & 72    & 72    & 72    & 72 \\
    \bottomrule
    \end{tabular}%
\begin{tablenotes}[flushleft]
\item Note: Both Model 1 and Model 2 use multilevel linear regressions with random intercepts for sessions and groups, while Model 2 further controls for the second-highest value in the group (V$_2$). Standard errors in the parentheses are adjusted for clustering at both session and group levels. *** Significant at the 1\% level, ** at the 5\% level, and * at the 10\% level.
\end{tablenotes}
\end{threeparttable}
\end{table}

\begin{result}
When the entry cost is low, Ind and Unr generate similar revenue, and they both generate significantly higher revenue than Res; when the entry cost is high, revenue generated by Ind is significantly higher than in Res and Unr. 
\end{result}
 
Having established the overall revenue rankings across mechanisms, we next disentangle the roles of entry decisions (\textit{participation effect}) and the selection process (\textit{selection effect}) in shaping revenue outcomes. 
Table~\ref{tab:two_effects} presents both observed and theoretically predicted effects, conditional on the realized values in each treatment.\footnote{These effects are slightly different from those in Table~\ref{tab: prediction}, which use the theoretical distribution of signals.}  To isolate these channels, we deliberately recalibrate the ``observed'' revenue rather than relying on raw experimental outcomes. First, we replace observed bids with equilibrium bids to eliminate the influence of over- or underbidding. Second, conditional on equilibrium bids and actual entry decisions, we compute expected ``observed'' revenue to abstract from random variation in bidder selection implemented in the experiment when the number of entrants exceeds the cap. 
Details of the recalibration are provided in Tables \ref{tab: Pred_given_signal} and \ref{tab:Observed_adjust_bidding} in Appendix~\ref{appendix:results}. 

\begin{center}
\begin{threeparttable}[h]
\footnotesize
\setstretch{1}
  \caption{\textit{Participation effect} and \textit{selection effect}}  \label{tab:two_effects}%
    \begin{tabular}{rccrccrccccc}
    \toprule
          & \multicolumn{5}{c}{\textbf{\textit{Participation effect}}} &       & \multicolumn{5}{c}{\textbf{\textit{Selection effect}}} \\
\cmidrule{1-6}\cmidrule{8-12}          & \multicolumn{2}{c}{\textbf{Low-cost}} &       & \multicolumn{2}{c}{\textbf{High-cost}} &       & \multicolumn{2}{c}{\textbf{Low-cost}} &       & \multicolumn{2}{c}{\textbf{High-cost}} \\
\cmidrule{1-3}\cmidrule{5-6}\cmidrule{8-9}\cmidrule{11-12}    

& \textbf{Obs} & \textbf{Pred} &       & \textbf{Obs} & \textbf{Pred} &       & \textbf{Obs} & \textbf{Pred} &       & \textbf{Obs} & \textbf{Pred} \\
 \multicolumn{1}{l}{Unr}   & 18.54 & 14.57 &       & 62.2 & 61.78 &       & 0     & 0     &       & 0     & 0 \\
          & \textcolor[rgb]{ .682,  .667,  .667}{(47.40)} & \textcolor[rgb]{ .682,  .667,  .667}{(45.04)} & \textcolor[rgb]{ .682,  .667,  .667}{} & \textcolor[rgb]{ .682,  .667,  .667}{(86.54)} & \textcolor[rgb]{ .682,  .667,  .667}{(83.89)} & \textcolor[rgb]{ .682,  .667,  .667}{} & \textcolor[rgb]{ .682,  .667,  .667}{(0)} & \textcolor[rgb]{ .682,  .667,  .667}{(0)} & \textcolor[rgb]{ .682,  .667,  .667}{} & \textcolor[rgb]{ .682,  .667,  .667}{(0)} & \textcolor[rgb]{ .682,  .667,  .667}{(0)} \\
       \multicolumn{1}{l}{Res}      & 14.93* & 10.82 &       & 48.63 & 47.45 &       & 16.18*** & 10.28 &       & 7.00*** & 2.04 \\
          & \textcolor[rgb]{ .682,  .667,  .667}{(46.77)} & \textcolor[rgb]{ .682,  .667,  .667}{(44.08)} & \textcolor[rgb]{ .682,  .667,  .667}{} & \textcolor[rgb]{ .682,  .667,  .667}{(73.15)} & \textcolor[rgb]{ .682,  .667,  .667}{(74.93)} & \textcolor[rgb]{ .682,  .667,  .667}{} & \textcolor[rgb]{ .682,  .667,  .667}{(14.89)} & \textcolor[rgb]{ .682,  .667,  .667}{(9.87)} & \textcolor[rgb]{ .682,  .667,  .667}{} & \textcolor[rgb]{ .682,  .667,  .667}{(12.20)} & \textcolor[rgb]{ .682,  .667,  .667}{(4.33)} \\
       \multicolumn{1}{l}{Ind}     & 10.65 & 9.50  &       & 33.07*** & 53.89 &       & 12.68*** & 4.83 &       & 7.55*** & 2.15 \\
          & \textcolor[rgb]{ .682,  .667,  .667}{(38.01)} & \textcolor[rgb]{ .682,  .667,  .667}{(39.85)} & \textcolor[rgb]{ .682,  .667,  .667}{} & \textcolor[rgb]{ .682,  .667,  .667}{(67.91)} & \textcolor[rgb]{ .682,  .667,  .667}{(82.40)} & \textcolor[rgb]{ .682,  .667,  .667}{} & \textcolor[rgb]{ .682,  .667,  .667}{(14.15)} & \textcolor[rgb]{ .682,  .667,  .667}{(6.24)} & \textcolor[rgb]{ .682,  .667,  .667}{} & \textcolor[rgb]{ .682,  .667,  .667}{(12.67)} & \textcolor[rgb]{ .682,  .667,  .667}{(4.60)} \\
    \bottomrule
    \end{tabular}%

\begin{tablenotes}[flushleft]
\item Note: Columns labeled "\textbf{Obs}" present the expected ``observed'' \textit{participation effect} and \textit{selection effect}, assuming participants follow equilibrium bids and given their entry choices. Columns "\textbf{Pred}" present the predicted effects, using the actual values drawn in the experiment and assuming participants all follow the equilibrium entry and bidding behaviors. Numbers in the round parentheses are the standard deviations. Asterisks next to the observed effects denote the significance level of the $t$-test (clustered at the group level) comparing the observed effects with the equilibrium predictions. *** Significant at the 1\% level, ** at the 5\% level, and * at the 10\% level.
\end{tablenotes}
\end{threeparttable}
\end{center}

\begin{center}

\begin{table}[htbp]
\begin{threeparttable}
  \footnotesize
  \caption{Multilevel linear regressions: \textit{participation effect} and \textit{selection effect}}
    \begin{tabular}{lccccrcccc}
    \toprule
          & \multicolumn{4}{c}{\textit{Participation effect}} &       & \multicolumn{4}{c}{\textit{Selection effect}} \\
\cmidrule{2-10}          & \multicolumn{2}{c}{Model 1} & \multicolumn{2}{c}{Model 2} &       & \multicolumn{2}{c}{Model 1} & \multicolumn{2}{c}{Model 2} \\
\cmidrule{2-5}\cmidrule{7-10}          & Low-cost & High-cost & Low-cost & High-cost &       & Low-cost & High-cost & Low-cost & High-cost  \\
\cmidrule{1-5}\cmidrule{7-10}    
    Unr (Cons.) & 18.54*** & 62.20*** & 16.94*** & 34.09*** & & -0    & -0    & -3.821** & -2.195* \\
          & (2.864) & (6.638) & (4.702) & (9.009) && (1.645) & (0.947) & (1.878) & (1.230) \\
    Res   & -3.617 & -13.57 & -3.600 & -11.98 && 16.18*** & 6.999*** & 16.22*** & 7.124*** \\
          & (4.050) & (9.388) & (4.041) & (9.458) && (2.326) & (1.340) & (2.298) & (1.333) \\
    Ind   & -7.892* & -29.13*** & -7.852* & -29.01*** && 12.68*** & 7.546*** & 12.77*** & 7.555*** \\
          & (4.050) & (9.388) & (4.042) & (9.452) && (2.326) & (1.340) & (2.298) & (1.332) \\
    V$_{2}$    &       &       & 0.010 & 0.172*** &&       &       & 0.023*** & 0.013*** \\
          &       &       & (0.022) & (0.037) &&       &       & (0.006) & (0.005) \\
    \# of Obs. & 720   & 720   & 720   & 720   && 720   & 720   & 720   & 720 \\
    \# of Groups & 72    & 72    & 72    & 72   & & 72    & 72    & 72    & 72 \\
    \bottomrule
    \end{tabular}%
\begin{tablenotes}[flushleft]
\item Note: Model 1 and Model 2 are both multilevel linear regressions with random intercepts for sessions and groups, while Model 2 further controls for the second-highest value in the group (V$_2$). Standard errors in the parentheses are adjusted for clustering at both session and group levels. *** Significant at the 1\% level, ** at the 5\% level, and * at the 10\% level.
\end{tablenotes}
  \label{tab:two_effects_reg}%
 \end{threeparttable}
\end{table}%
    
\end{center}

As shown in the summary statistics given in Table \ref{tab:two_effects} (left panel), the observed \textit{participation} effects are generally slightly higher than predicted--though not statistically significantly so--across both cost conditions, with only one exception.  In the \textit{Ind} treatment under high entry cost, the observed \textit{participation effect} is substantially smaller than predicted (33.07 vs 53.89, $p < 0.01$). 

To further examine \textit{participation} effects across mechanisms, Table \ref{tab:two_effects_reg} reports multilevel linear regressions of \textit{participation} and \textit{selection} effects controlling for both session and group level random effects.\footnote{Similar to revenue comparisons, we also control for the second-highest values in Model 2 and the results are qualitatively similar and our main discussion focuses on the results of Model 2.} Consistent with theoretical predictions, \textit{Ind} exhibits a smaller \textit{participation effect} than \textit{Unr} under both cost conditions (weakly significant under \textit{low-cost}, $p = 0.052$; strongly significant under \textit{high-cost},  $p < 0.01$). While the estimated \textit{participation} effects are also smaller in \textit{Res} than in \textit{Unr}, which aligns with the theory, the differences are not statistically significant. Similarly, the difference between \textit{Ind} and \textit{Res} aligns with the direction of the theoretical prediction but is not statistically significant ($p = 0.29$) under the \textit{low-cost} condition and is only marginally significant under the \textit{high-cost} condition ($p = 0.07$).

\begin{result}

The observed \textit{participation} effects are largely consistent with theoretical predictions. Two deviations arise: the participation effect is significantly smaller than predicted in \textit{Ind} under high entry cost and is marginally larger than predicted in \textit{Res} under low entry cost. As predicted, the \textit{participation effect} is lower in \textit{Ind} than in \textit{Unr} under both cost conditions, whereas no statistically significant difference emerges between \textit{Res} and \textit{Unr}, though the sign of these differences aligns with the theory.
 \end{result}

We next turn to the \textit{selection effect}. As shown in the right panel of Table \ref{tab:two_effects}, observed \textit{selection} effects in both \textit{Res} and \textit{Ind} are significantly larger than predicted, indicating greater selection inefficiency than implied by theory.\footnote{No selection occurs in \textit{Unr} and both predicted and observed \textit{selection} effects are therefore zero.} This inefficiency reduces revenue in both mechanisms more than expected. Under high entry cost, theory predicts that the additional message in \textit{Ind} is not useful for separating bidders with high signals from those with low signals, implying similar \textit{selection} effects in \textit{Res} and \textit{Ind}. This prediction is supported by the data: observed selection effects are statistically indistinguishable ($p = 0.75$) based on Model 2 in Table \ref{tab:two_effects_reg}. Under low entry cost, the additional message in \textit{Ind} is predicted to improve selection by allowing high-signal bidders to separate more effectively. While the observed \textit{selection effect} is indeed smaller in \textit{Ind} than in \textit{Res}, the difference is not statistically significant ($p = 0.13$).

\begin{result}
In \textit{Ind} and \textit{Res}, \textit{selection} effects are consistently larger than predicted by theory. Selection inefficiency affects \textit{Ind} and \textit{Res} similarly under both \textit{low-cost} and \textit{high-cost} conditions.
\end{result}

Taken together, these findings explain the observed revenue ranking. Under high entry cost, \textit{Ind} generates substantially higher revenue than the other mechanisms, driven primarily by greater-than-predicted participation and a correspondingly smaller \textit{participation effect}. Under low entry cost, the revenue advantage of \textit{Ind} relative to \textit{Unr} is offset by a larger-than-predicted \textit{selection effect}, eliminating the expected revenue gain. Although \textit{Res} is predicted to perform well under high entry cost, it generates the lowest revenue under both cost conditions, largely due to a larger-than-expected \textit{participation effect} in the \textit{low-cost} condition and a persistently larger-than-expected \textit{selection effect} under both cost conditions.

\subsection{Individual behavior}\label{subsec:ind choices}
Because the treatment-level revenue patterns and deviations from theoretical predictions arise from individual decisions, we now turn to participants' entry and bidding behavior to account for these results.

\subsubsection{Bidding Behavior}\label{subsec:bidding}

\begin{figure}[h!]
    \centering
    \includegraphics[width=0.8\textwidth]{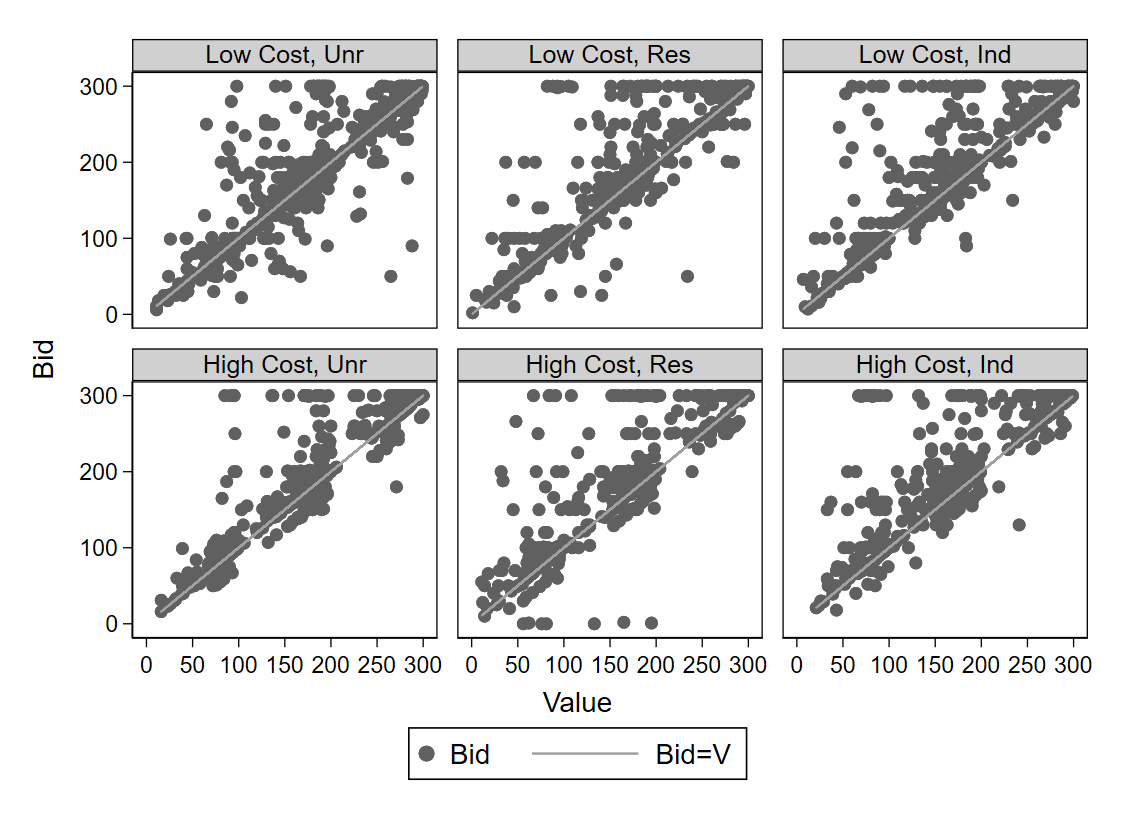}
    \caption{Bidding behavior}
    \label{fig:bid}
\end{figure}

Theory predicts that in a second-price auction, bidders should bid their true valuations. Figure \ref{fig:bid} plots bids (vertical axis) against valuations (horizontal axis). Most observations lie on or near the 45-degree line. Among deviations, overbidding appears to be more prevalent than underbidding across all treatments, a pattern well documented in the experimental literature \citep{kagel2016auctions}. Within each treatment, both over- and underbidding decline as the entry cost increases, with a larger reduction in underbidding. Because \textit{high-cost} rounds follow \textit{low-cost} rounds, this pattern suggests that overbidding remains relatively persistent in second-price auctions.

To quantify bidding behavior, we compute the bid-value ratio (\textit{B-V ratio}), defined as the observed bid divided by the bidder's valuation. A \textit{B-V ratio} greater (less) than one indicates over- (under-) bidding. We regress the \textit{B-V ratio} on treatment indicators separately for the \textit{low-cost} and \textit{high-cost} conditions using a multilevel linear model. Table \ref{tab:bid} reports the results.

When the entry cost is low, average bids exceed valuations by 15\%, 22\%, and 35\% in \textit{Unr}, \textit{Res}, and \textit{Ind}, respectively ($p=0.04$, $p < 0.01$ and $p < 0.01$). When the entry cost is high, overbidding remains substantial at 17\%, 29\%, and 29\%. Across treatments within a given cost condition, differences in overbidding are generally not statistically significant; only the difference between \textit{Ind} and \textit{Unr} in the \textit{low-cost} condition is marginally significant at the 10\% level ($p = 0.08$).

The similarity in bidding behavior across treatments suggests that revenue differences are not driven by bidding strategies. To verify this, we re-estimate the multilevel linear regressions in Table \ref{tab: rev} using counterfactual revenues constructed from observed entry decisions and equilibrium bidding strategies (see Table \ref{tab:rev_reg_hypothetical} in Appendix \ref{appendix:results}). The estimated treatment effects and their statistical significance remain unchanged.

\begin{table}[h!]
\centering
  \footnotesize
  \caption{Multilevel linear regressions on \textit{B-V ratio}}
    \begin{threeparttable}
  \centering
    \begin{tabular}{lccccc}
    \toprule
          & \multicolumn{2}{c}{Low-cost} &       & \multicolumn{2}{c}{High-cost} \\
    \midrule
          & Coef  & \textit{Wald} test & & Coef  & \textit{Wald} test \\
\cmidrule{2-3}\cmidrule{5-6}    Unr (Cons.) & 1.15*** & $p=0.04$ &       & 1.17*** & $p < 0.01$ \\
          & (0.08) &       &       & (0.06) &  \\
    Res    & 0.07 & $p < 0.01$ &       & 0.12 & $p<0.01$ \\
          & (0.11) &       &       & (0.09) &  \\
    Ind   & 0.20* & $p < 0.01$ &       & 0.12 & $p < 0.01$ \\
          & (0.11) &       &       & (0.09) &  \\
$\sigma^2_{group}$&0.08&&& 0.02& \\
                  & (0.03)&&&(0.02)&\\
$\sigma^2_{individual}$&0&&&0.07&\\
                    & (0)&&&(0.03)&\\
\# of Obs. & 1,581 &       &       & 1,252 &  \\
\# of Individuals & 352    &       &       & 331    &  \\
    \bottomrule
\end{tabular}%
    \begin{tablenotes}[flushleft]
    \item \noindent Note: Multilevel linear regressions control for both group and individual levels of random effects. The number of observations (\# of Obs.) is the number of observed bids conditional on entry. The number of individuals (\# of Individuals) is the number of individual-level clusters in the multilevel regression. \textit{Wald} tests are used to compare the observed \textit{B-V ratio} with 1.*** Significant at the 1\% level, ** at the 5\% level, and * at the 10\% level.
    \end{tablenotes}
    \end{threeparttable}
  \label{tab:bid}%
\end{table}%

\begin{result}
Consistent with prior experimental evidence on second-price auctions, bidders systematically overbid in all treatments. Overbidding rates do not differ significantly across treatments and do not drive treatment-level revenue differences.   
\end{result}

\subsubsection{Entry behavior} \label{sec: entry_be}
\begin{figure}[ht] 
 \caption{Entry choices compared to equilibrium predictions}
\begin{center}
\includegraphics[width=0.8\textwidth]{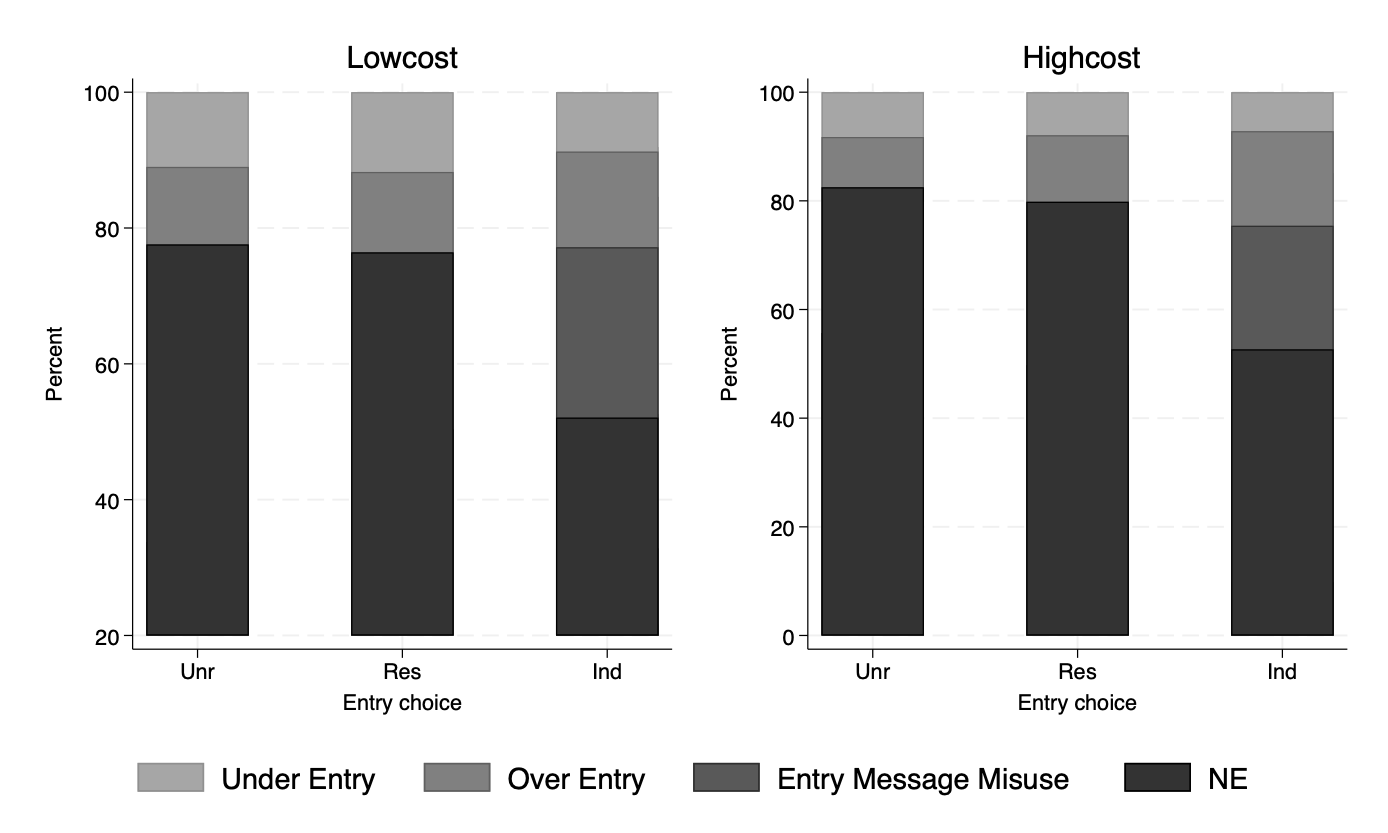}
\label{fig: entry_break} 
\end{center}  
\noindent\footnotesize Note: ``Under Entry" refers to cases in which the chosen message is $m=0$ when the equilibrium predicts $m=1$ or $m=2$. ``Over Entry" refers to cases in which the chosen message is $m=1$, or $m=2$ when the equilibrium predicts $m=0$. ``Entry Message Misuse'' captures cases in which the positive message chosen does not match the predicted positive entry message: specifically, $m=1$ is chosen when the equilibrium predicts $m=2$, or vice versa. ``NE" denotes cases in which the chosen message coincides with the equilibrium prediction. 
\end{figure}

Turning to entry decisions, deviations from equilibrium entry behavior amplify the \textit{selection effect} in both the \textit{Res} and \textit{Ind} treatments relative to theoretical predictions. We classify participants' entry choices into four categories: Nash entry, in which the chosen entry message coincides with the equilibrium message given the initial signal; over-entry, in which the chosen message is positive while the equilibrium prediction is zero; under-entry, in which the chosen message is zero while the equilibrium message is positive; and (positive) entry message misuse, in which the chosen message is ``1" when the equilibrium is ``2" or vice versa. As shown in Figure \ref{fig: entry_break}, patterns of over- and under-entry are stable across cost conditions within each mechanism and are qualitatively similar between the \textit{Unr} and \textit{Res} treatments. Under the \textit{low-cost} condition, Nash entry accounts for 78\% (\textit{Unr}) and 76\% (\textit{Res}) of decisions; under the \textit{high-cost} condition, these shares increase slightly to 83\% and 80\%, respectively. The remaining choices are approximately evenly split between over- and under-entry.\footnote {In the \textit{low-cost} condition, over-entry and under-entry each account for 11\% of decisions in \textit{Unr}, and 12\% in \textit{Res}. In the \textit{high-cost} condition, under-entry accounts for 8\% of decisions in both \textit{Unr} and \textit{Res}, while over-entry accounts for 9\% in \textit{Unr} and 12\% in \textit{Res}. Table \ref{tab:entry_breakdown} in Appendix \ref{appendix:results} provides the full breakdown.} As a result, the distributions of realized number of entrants in \textit{Unr} and \textit{Res} closely track their equilibrium counterparts under both cost conditions (see Figure \ref{fig:freq_n} in Appendix \ref{appendix:results}). 

In contrast, only 52\% of entry choices under the \textit{low-cost} condition and 53\% under the \textit{high-cost} condition in the \textit{Ind} treatment are consistent with equilibrium predictions. In both cost conditions, over-entry is slightly more prevalent than under-entry: 14\% vs.\ 9\% under \textit{low-cost} and 17\% vs.\ 7\% under \textit{high-cost}. The greater complexity of the entry decision in \textit{Ind} plausibly reduces equilibrium-consistent behavior, which in turn amplifies the \textit{selection effect} relative to theoretical predictions. Moreover, an additional 25\% (\textit{low-cost}) and 23\% (\textit{high-cost}) of entry choices involve misuse of positive entry messages, predominantly cases in which participants choose $m=2$ when the equilibrium predicts $m=1$ (the reverse occurs in only 0.5\% of cases under the \textit{low-cost} condition). The systematic overuse of the highest entry messages is the main source of selection inefficiency observed in the \textit{Ind} treatment.

Because a second-price auction without a reserve price generates zero revenue when fewer than two bidders enter, over-entry increases revenue primarily by converting auctions with zero or one entrant into auctions with at least two entrants. In the \textit{Ind} treatment under the \textit{high-cost} condition--where theory predicts a substantial share of non-competitive auctions (see Figure~\ref{fig:freq_n} in Appendix \ref{appendix:results})--over-entry reduces the fraction of such auctions from 37.09\% (predicted) to 19.58\% (observed). This shift attenuates the \emph{participation effect} relative to theoretical predictions and pushes realized revenue above the predicted level. In contrast, under the \textit{low-cost} condition, the predicted share of non-competitive auctions is already small (approximately 7\%), leaving limited scope for over-entry to further reduce the \textit{participation effect} or increase revenue.

To formally compare entry behavior across treatments, we estimate the probabilities of each message being chosen conditional on the initial signal $S_i$ (see Table \ref{tab: p_entry}). Across both cost conditions, participants in the \textit{Ind} treatment are significantly more likely to choose positive messages ($m\in\{1,2\}$) than those in \textit{Res} ($m=1$), while participants in \textit{Res} are more likely to opt in than those in \textit{Unr} (all $p<0.01$).

\begin{result}
Approximately 80\% of entry choices in the \textit{Res} and \textit{Unr} treatments align with equilibrium predictions under both cost conditions, with the remaining choices roughly evenly split between over- and under-entry. In contrast, only about 50\% of the entry choices in the \textit{Ind} treatment are consistent with equilibrium, primarily due to slightly higher over-entry and a substantial share of misuse of the highest entry message. Conditional on the initial signal, willingness to opt in is the highest in \textit{Ind}, followed by \textit{Res}, and lowest in \textit{Unr}. 
\end{result}

\begin{center}
\begin{threeparttable} 
\setstretch{1}
\footnotesize
\caption{Estimated average probability of the entry message being chosen}
\begin{tabular}{rccrcccccccc}
\toprule
& \multicolumn{3}{c}{\textbf{Unr}} && \multicolumn{3}{c}{\textbf{Res}} && \multicolumn{3}{c}{\textbf{Ind}} \\
\midrule
& \textbf{NE}& \textbf{Mean} & \textbf{$p$} && \textbf{NE} & \textbf{Mean} & \textbf{$p$} && \textbf{NE} & \textbf{Mean} & \textbf{$p$} \\
\cmidrule{2-4}\cmidrule{6-8}\cmidrule{10-12}\multicolumn{1}{l}{\textbf{Low-cost}} &&&&&&&&&&&  \\
\textbf{m=1} & 0.55 & 0.55& 0.73&& 0.62& 0.64 & 0.10 && 0.44  & 0.22  & 0.00 \\
 && (0.372) && && (0.352) &&&& (0.160) &  \\
\textbf{m=2} &&&&&&&&& 0.20  & 0.51  & 0.00 \\
 &&&&&&&&&& (0.376) &  \\
\multicolumn{1}{l}{\textbf{High-cost}} &&&&&&&&&&&  \\
\textbf{m=1}& 0.35& 0.36 & 0.46 && 0.38 & 0.42 & 0.00 &  & 0.40  & 0.22  & 0.00 \\
 && (0.352) &&&& (0.369) &&&& (0.157) &  \\
\textbf{m=2} &&&&&&&&& 0& 0.29  & 0.00 \\
 &&&&&&&&&& (0.324) &  \\
    \bottomrule
    \end{tabular}%
  \label{tab: p_entry}
\begin{tablenotes}[flushleft]
\item Note: The probability of each message being chosen, conditional on  $S_i$, is computed using the estimated cutoff points and coefficients of the multilevel ordered logistic model with random effects at both the individual and group levels. The multilevel ordered logistic regression results are presented in Table \ref{tab:entrychoice_melogit} in Appendix C. We omit the estimated probability of $m=0$, as it can be derived from $1-p(m>0)$. $P$-values are based on \textit{Wald} tests clustered at the group level, comparing the observed probabilities with the corresponding predicted entry thresholds. Standard deviations are reported in parentheses.
\end{tablenotes}
\end{threeparttable}
\end{center}

\subsection{Social welfare}\label{subsec:profit}

To assess the overall impact of the mechanisms on both sellers and bidders, we compare bidders' profits and social welfare--defined as the sum of bidders' profits and seller's revenue--across different treatments. Bidders' profit is calculated as the sum of profits earned by all five potential bidders in each group in each round.\footnote{See Tables \ref{tab:bid_profit_sum} and \ref{tab:social_walfare_sum} in Appendix \ref{appendix:results} for summary statistics.}

\begin{table}[htbp]
\begin{center}
\begin{threeparttable}
\footnotesize
  \caption{Social welfare and bidders' profit: Multilevel linear regressions}
    \begin{tabular}{lccccrcccc}
    \toprule
          & \multicolumn{4}{c}{Social welfare} &       & \multicolumn{4}{c}{Bidders' profit} \\
\cmidrule{2-10}          & \multicolumn{2}{c}{Model 1} & \multicolumn{2}{c}{Model 2} &       & \multicolumn{2}{c}{Model 1} & \multicolumn{2}{c}{Model 2} \\
          & Low-cost & High-cost & Low-cost & High-cost &       & Low-cost & High-cost & Low-cost & High-cost \\
\cmidrule{2-5}\cmidrule{7-10}    
Unr(Cons.) & 157.2*** & 114.1*** & -27.14*** & -59.08*** &       & 0.962 & 7.025 & -13.36*** & -26.65*** \\
          & (5.320) & (5.394) & (3.285) & (4.374) &       & (3.752) & (8.890) & (4.360) & (9.434) \\
    Res    & 0.304 & 2.154 & 2.038 & 12.06*** &       & 15.81*** & 4.275 & 15.84*** & 6.365 \\
          & (7.523) & (7.628) & (2.929) & (4.230) &       & (5.307) & (12.57) & (5.177) & (11.81) \\
    Ind   & 3.083 & 9.479 & 5.397* & 11.39***&       & 10.30* & -15.31 & 8.623* & -12.94 \\
          & (7.523) & (7.628) & (2.929) & (4.226) &       & (5.307) & (12.57) & (5.184) & (11.81) \\
       $V_1$  &       &       & 1.001*** & 0.950*** &       &       &       &       &  \\
          &       &       & (0.0138) & (0.0175) &       &       &       &       &  \\
    $V_1-V_2$ &       &       &       &       &       &       &       & 0.909*** & 1.792*** \\
          &       &       &       &       &       &       &       & (0.150) & (0.233) \\
 
    Obs. & 720   & 720   & 720   & 720   &       & 720   & 720   & 720   & 720 \\
    Groups & 72     & 72     & 72     & 72     &       & 72    & 72    & 72    & 72 \\
    \bottomrule
    \end{tabular}%
  \label{tab:reg_welfare}%
\begin{tablenotes}[flushleft]
\item Note:  Model 1 and Model 2 are both multilevel linear regressions with random intercepts for sessions and groups, while Model 2 further controls for the highest value in the group ($V_1$) in social welfare and controls for the difference between the highest value and the second-highest value in the group ($V_1-V_2$) in bidders' profit. Standard errors in parentheses are adjusted for clustering at both session and group levels. *** Significant at the 1\% level, ** at the 5\% level, and * at the 10\% level.
\end{tablenotes}
 \end{threeparttable}
 \end{center}
\end{table}%

The left panel of Table \ref{tab:reg_welfare} reports multilevel linear regressions of social welfare on treatment indicators for each cost condition.\footnote{Model 2 includes a control variable for the highest realized value $V_1$ in regressions on social welfare and for the difference between the highest and second-highest realized values in regressions on bidders' profit (right panel of Table \ref{tab:reg_welfare}). The social welfare generated by an auction in the complete information setting is equal to $V_1$--the sum of auction revenue $V_2$ and the winning bidder's profit $V_1-V_2$. Controlling for the drawn values of $V_1$ on social welfare and for $V_1-V_2$ on bidders' profit eliminates variations caused by random draws across treatments. We only discuss results from model 2, given that all models give qualitatively similar results.} Under both cost conditions, social welfare is higher in \textit{Ind} than in \textit{Unr} ($p=0.065$ and $p  < 0.01$, respectively). In contrast, the difference between \textit{Ind} and \textit{Res} is not statistically significant under either cost condition ($p=0.252$ and $p=0.876$, respectively). Comparing \textit{Res} and \textit{Unr}, \textit{Res} generates significantly higher social welfare only when the entry cost is high ($p < 0.01$).

The right panel of Table \ref{tab:reg_welfare} reports regressions on bidders' profits. When the entry cost is low, bidders earn higher profits in both \textit{Res} and \textit{Ind} (by 8.62 and 15.84 units) relative to  \textit{Unr} ($p=0.096$ and $p=0.002$), while profits do not differ significantly between \textit{Res} and \textit{Ind} ($p=0.164$). When the entry cost is high, bidders' profits do not differ significantly between \textit{Res} and \textit{Unr} ($p = 0.590$) or between \textit{Ind} and \textit{Unr} ($ p = 0.273$). Bidders earn around 19 units lower profits in \textit{Ind} than in \textit{Res}, although this difference is not statistically significant ($p=0.102$). 
While the higher-than-predicted participation in \textit{Ind} raises seller revenue compared to \textit{Res}, lower bidders' profits lead to no statistically significant difference in social welfare between \textit{Ind} and \textit{Res}.

\begin{result}
Relative to \textit{Unr}, social welfare is higher in \textit{Ind} for both cost conditions and higher in \textit{Res} under the high-cost condition. Social welfare does not differ significantly between \textit{Ind} and \textit{Res} under either cost condition. Bidders earn higher profits in \textit{Ind} and \textit{Res} than in \textit{Unr} under the low-cost condition, but earn similar profits across the three treatments under the high-cost condition.
\end{result}

\section{Conclusion} \label{sec:conclusion}
This paper combines a controlled laboratory experiment with a supporting theoretical framework to study how entry restrictions and indicative bidding affect revenue and welfare in auctions with costly entry. We show that restricting entry involves a fundamental trade-off: while it encourages participation by reducing expected competition conditional on entry, it may generate inefficiencies through imperfect selection. By formally defining and empirically measuring the \textit{participation} and \textit{selection} effects, we quantify the relative strength of these opposing forces.

Our theoretical analysis predicts that imposing an entry cap can increase seller revenue when entry costs are sufficiently high, extending \cite{moreno2011auctions} to environments in which bidders receive informative signals before entry. We further show that, for some environments, indicative bidding with additional entry messages can increase seller revenue relative to both a simple entry cap and unrestricted entry.

The experimental results broadly support these predictions, but also reveal systematic deviations from equilibrium behavior. In particular, off-equilibrium entry decisions generate larger-than-predicted selection inefficiencies, preventing restricted entry from outperforming unrestricted entry when entry costs are high. Indicative bidding consistently increases seller revenue relative to unrestricted entry; however, contrary to theoretical predictions, these gains arise primarily from increased participation rather than improved selection.

These findings suggest that while indicative bidding can be effective in practice, its performance depends critically on potential bidders’ ability to sort by expected values at the entry stage--a goal shared by both sellers and bidders. One potential avenue for improving selection is the use of sequential or multi-step screening mechanisms rather than a single, simultaneous message choice \citep{sweeting2015selective, lu2021orchestrating}. Designing such mechanisms, however, may impose additional strategic complexity on bidders. Future experimental work could examine how sequential screening affects participation, selection, and welfare in environments with costly information acquisition.

\newpage

\bibliography{ref}
\newpage

\begin{appendices}
\section{Equations and Proofs}\label{appendix:proofs}

\subsection*{Equations}

Given an entry mechanism with $\overline M=1$, the expected revenue of the auction is
\begin{align}
    \Pi(\mathcal{M},\mathcal{X}) & = \sum_{k=2}^N q_k \left(\alpha_0 + \frac{\min\{k,n\} - 1}{\min\{k,n\} + 1}(\bar S - \alpha_0) + \mathbb{E}[T]\right),\label{equation:rev m=1}
\end{align}
where $\alpha_0$ is determined by (\ref{RES cut-off}) and $q_k = {N \choose k} \left(\frac{\bar S -\alpha_0}{\bar S} \right)^k \left(\frac{\alpha_0}{\bar S} \right)^{N-k}$ is the probability that $k$ potential bidders choose $m=1$. The expected revenue of the counterfactual scenario is
\begin{align}
    \hat\Pi(\alpha_0(\mathcal{M}),\mathcal{X}) & = \sum_{k=2}^N q_k \left(\alpha_0 + \frac{k - 1}{k + 1}(\bar S - \alpha_0) + \mathbb{E}[T]\right),\label{equation:cf rev m=1}
\end{align}
and the \textit{participation} and \textit{selection effects} are given by
\begin{align} 
PE(\mathcal{M},\mathcal{X}) & = \frac{N-1}{N+1} \bar S + \mathbb{E}[T] - \sum_{k=2}^N q_k\left(\frac{2\alpha_0}{k+1} + \frac{k-1}{k+1} \bar S + \mathbb{E}[T]\right), \text{ and}\label{equation:PE}\\
SE(\mathcal{M},\mathcal{X}) & = \sum_{k=n+1}^N q_k \left(\frac{k - 1}{k + 1} - \frac{n - 1}{n + 1}\right)(\bar S - \alpha_0).\label{equation:SE}
\end{align}

For an entry mechanism with $\overline M = 2$ the revenue is
\begin{align}
    \Pi(\mathcal{M},\mathcal{X}) & = \sum_{k=2}^N q_{k,0} \left(\alpha_0 + \frac{\min\{k,n\} - 1}{\min\{k,n\} + 1}(\alpha_1 - \alpha_0) + \mathbb{E}[T]\right)\nonumber\\
    & + \sum_{k=1}^{N-1} q_{k,1} \left(\alpha_0 + \frac{\min\{k,n-1\}}{\min\{k,n-1\} + 1}(\alpha_1 - \alpha_0) + \mathbb{E}[T]\right)\nonumber\\
    & + \sum_{j=2}^N \sum_{k=0}^{N-j} q_{k,j} \left(\alpha_1 + \frac{\min\{j,n\} - 1}{\min\{j,n\} + 1}(\bar S - \alpha_1) + \mathbb{E}[T]\right),\label{equation:rev ind}
\end{align}
where $q_{k,j} = {N \choose j} {N-j \choose k} \left(\frac{\bar S - \alpha_1}{\bar S}\right)^j \left(\frac{\alpha_1 -\alpha_0}{\bar S} \right)^k \left(\frac{\alpha_0}{\bar S} \right)^{N-j-k}$ is the probability that $j$ bidders choose $m=2$ and $k$ others choose $m=1$.

The revenue of the counterfactual situation and the \textit{participation effect} are the same as in the $\overline M = 1$ case.

The \textit{selection effect} with $\overline M=2$ is given by
\begin{align}
    SE(\mathcal{M},\mathcal{X}) = \sum_{k=n+1}^N & q_{k,0} \left(\frac{k - 1}{k + 1} - \frac{n - 1}{n + 1}\right)(\alpha_1 - \alpha_0) +  \sum_{k=n}^{N-1} q_{k,1} \left(\frac{k}{k + 1} - \frac{n - 1}{n}\right)(\alpha_1 - \alpha_0)\nonumber\\
    & + \sum_{j=n}^N \sum_{k=0}^{N-j} q_{k,j} \left(\frac{j - 1}{j + 1} - \frac{n - 1}{n + 1}\right)(\bar S - \alpha_1) \label{equation:se ind}
\end{align}
where $q_{k,j} = {N \choose j} {N-j \choose k} \left(\frac{\bar S - \alpha_1}{\bar S} \right)^j \left(\frac{\alpha_1 -\alpha_0}{\bar S} \right)^k \left(\frac{\alpha_0}{\bar S} \right)^{N-j-k}$.

Given an entry mechanism with $\overline M=1$, the expected aggregate profit of the bidders is
\begin{align}
    Y(\mathcal{M},\mathcal{X}) & = \sum_{k=2}^N q_k \left(\frac{(\bar S - \alpha_0)}{\min\{k,n\}+1} - \min\{k,n\} c\right) + q_1\left(\frac{\bar S + \alpha_0}{2} - c +\mathbb{E}[T]\right).\label{equation:profit res}
\end{align}
For $\overline M=2$ the aggregate profit of the bidders is
\begin{align}
   & Y(\mathcal{M},\mathcal{X}) = \sum_{k=2}^N q_{k,0} \left(\frac{\alpha_1 - \alpha_0}{\min\{k,n\}+1} - \min\{k,n\} c \right) + \sum_{k=1}^{N-1} q_{k,1} \left(\frac{\bar S - \alpha_0}{\min\{k,n\}} - \min\{k,n\}c \right)\nonumber\\
   \ \  & + \sum_{j=2}^N \sum_{k=0}^{N-j} q_{k,j} \left(\frac{\bar S - \alpha_1}{\min\{j,n\}+1} - \min\{j,n\}c \right) + (q_{0,1} + q_{1,0}) \left( \frac{\bar S + \alpha_0}{2} - c +\mathbb{E}[T]\right) .\label{equation:profit ind}
\end{align}

\subsection*{Proof of Lemma~\ref{lemma:entry restriction cutoff}}

Given $\overline M = 1$, for any $n=2,\ldots,N$ the cutoff signal $\alpha_0$ is uniquely determined by $(\ref{RES cut-off})$. Uniqueness follows from arguments provided in the online appendix of \cite{quint2018theory} that hold for the environments used here. The LHS of this equation is continuous in $\alpha_0$, is negative when $\alpha_0 = 0$ and is positive for $\alpha_0 = \bar S$ whenever $c < \bar S + \mathbb{E}[T]$. Then the LHS must cross zero exactly once from below. For $n',n\in\{2,\ldots,N\}$ and $n'>n$ the LHS is strictly smaller for $n'$ than for $n$ for every $\alpha_0 \in(0,\bar S)$. Then the LHS must cross zero at an $\alpha_0'>\alpha_0$.

\subsection*{Proof of Proposition~\ref{proposition:entry restriction}}

From Lemma~\ref{lemma:entry restriction cutoff}, for each $n\in\{2,\ldots,N\}$ there is a unique cutoff signal $\alpha_0(n)$ where $\alpha_0(n') > \alpha_0(n)$ for $n'>n$. Because the equilibrium messaging strategy is a cutoff strategy and $T$ is common to all potential bidders' values, the \textit{participation effect} equals the expected revenue loss from the event that the second-highest-valuation bidder opts out. Conditional on this event for a given signal profile the loss is $S_{(2)} + \mathbb{E}[T]$, where $S_{(2)}$ is the second-highest signal.

Fix $n < n'$ and consider the event $S_{(2)}\in(\alpha_0(n),\alpha_0(n'))$. In this event, the second-highest-signal bidder opts in when the cap is $n$, but opts out when the cap is $n'$. Thus the dropout occurs under $n'$ but not under $n$, implying a larger participation effect under $n'$. In the complement event, the second-highest-signal bidder makes the same entry decision under both caps, so the \textit{participation effect} is unchanged. Therefore, the \textit{participation effect} is higher under $n'$ than $n$.

From (\ref{equation:SE}), the \textit{selection effect} can be written as  $SE(\mathcal{M},\mathcal{X}) = \sum_{k=0}^N q_k f(n,\alpha_0;k)$ where $f(n,\alpha_0;k)=0$ for $k=0,\ldots,n$ and $f(n,\alpha_0;k) = \left(\frac{k-1}{k+1} - \frac{n-1}{n+1}\right)(\bar S - \alpha_0)$ for $k=n+1,\ldots,N$. From Lemma~\ref{lemma:entry restriction cutoff}, $\alpha_0(n') > \alpha_0(n)$, hence $f(n',\alpha_0(n');k) \le f(n,\alpha_0(n);k)$ for all $k$, with strict inequality for $k\ge n+1$. 

Recall $q_k(n)=\Pr(K=k)$ for $K\sim\text{Bin}(N,p(n))$ where 
$p(n)=\frac{\bar S-\alpha_0(n)}{\bar S}$. For $n'>n$, $p(n')<p(n)$, so $K'\sim\text{Bin}(N,p(n'))$ is first-order stochastically dominated by $K$. Since $f(n',\alpha_0(n');k)$ is weakly increasing in $k$, we obtain
\[
\sum_{k=0}^N q_k(n') f(n',\alpha_0(n');k)
\le \sum_{k=0}^N q_k(n) f(n',\alpha_0(n');k) < \sum_{k=0}^N q_k(n) f(n,\alpha_0(n);k).
\]
Hence, the \textit{selection effect} is lower under $n'$ than under $n$.

\subsection*{Proof of Proposition~\ref{proposition:PE vs SE}}
(a) Given that the total revenue loss is the sum of the \textit{participation effect} and the \textit{selection effect}, to show the result we will show that for sufficiently small $c$, the difference $SE((n,\overline M=1),\mathcal{X}) - SE((N,\overline M=1),\mathcal{X})$ is larger than $PE((n,\overline M=1),\mathcal{X}) - PE((N,\overline M=1),\mathcal{X})$ for all $n \in \{2,\ldots,N-1\}$. Given that $SE((N,\overline M=1),\mathcal{X})=0$, the difference in \textit{selection effect} is equal to $SE((n,\overline M=1),\mathcal{X})$, which from (\ref{equation:SE}) is
\[
SE((n,\overline M=1),\mathcal{X}) - SE((N,\overline M=1),\mathcal{X}) = \sum_{k=n+1}^N q_k \left(\frac{k-1}{k+1} - \frac{n-1}{n+1} \right) (\bar S - \alpha_0),
\]
where $\alpha_0$ and therefore $q_k$ depend on both $n$ and $c$. For any $n\in \{2,\ldots,N-1\}$, $\alpha_0\to 0$ as $c\to 0$. Then $q_k = {N \choose k} \left(\frac{\bar S -\alpha_0}{\bar S} \right)^k \left(\frac{\alpha_0}{\bar S} \right)^{N-k} \to 0$ for $k<N$ and $q_N \to 1$. Then the difference in \textit{selection effect} continuously approaches $\left(\frac{N-1}{N+1} - \frac{n-1}{n+1} \right) \bar S$ which is strictly positive for all $n<N$.

From (\ref{equation:PE}) as $c\to0$ and therefore $\alpha_0\to0$ and $q_k\to 0$ for $k<N$ and $q_N\to 1$, it follows that for $n\in\{2,\ldots,N\}$, $PE((n,\overline M=1),\mathcal{X})\to 0$ as $c\to 0$. Then the difference in \textit{participation effect} approaches zero and is less than the difference in \textit{selection effect} for all $n\in\{2,\ldots,N - 1\}$.\\

(b) For a given $\varepsilon>0$, let $c = \bar S + \mathbb{E}[T] - \varepsilon$ and $\alpha_0(\varepsilon) = \bar S - \delta(\varepsilon)$. Substituting these into (\ref{RES cut-off}) we get
\begin{equation} 
 p_0 (\varepsilon - \delta(\varepsilon)) - c \sum_{k=1}^{N-1} p_k \min\left\{1,\frac{n}{k+1}\right\} = 0. \label{equation:RES cut-off varepsilon}
\end{equation}
For $\varepsilon < \bar S + \mathbb{E}[T]$, $c>0$ and therefore the first term of (\ref{equation:RES cut-off varepsilon}) is positive, which implies that $\delta(\varepsilon)\in(0,\varepsilon)$.
It follows that as $c\to \bar S + \mathbb{E}[T]$, both $\varepsilon\to 0$ and $\delta(\varepsilon) \to 0$.

Using this we quantify the impact of changing $n$ on $\delta(\varepsilon)$ for small $\varepsilon$. We can approximate the cutoff condition for $\delta(\varepsilon)$ close to zero using a Taylor expansion of the $p_k$ terms. Below we highlight terms of an order greater than or equal to $\delta^2(\varepsilon)$.
\begin{align*}
p_0 & = \left(\frac{\alpha_0}{\bar{S}}\right)^{N-1} = \left(1 - \frac{\delta(\varepsilon)}{\bar S} \right)^{N-1} = 1 - \frac{(N - 1) \delta(\varepsilon)}{\bar S} + \frac{(N - 1)(N-2) \delta^2(\varepsilon)}{2 \bar S^2} + \smallO(\delta^2(\varepsilon))\\
p_1 & = \binom{N-1}{1}\left(\frac{\delta(\varepsilon)}{\bar S}\right)\left(1 - \frac{\delta(\varepsilon)}{\bar S} \right)^{N-2} = \frac{(N-1)\delta(\varepsilon)}{\bar S} - \frac{(N-1)(N-2)\delta^2(\varepsilon)}{\bar S^2} + \smallO(\delta^2(\varepsilon))\\
p_2 & = \binom{N-1}{2}\left(\frac{\delta(\varepsilon)}{\bar S}\right)^2\left(1 - \frac{\delta(\varepsilon)}{\bar S} \right)^{N-3} = \frac{(N-1)(N-2)\delta^2(\varepsilon)}{2\bar S^2} + \smallO(\delta^2(\varepsilon))
\end{align*}
Any $p_k$ for $k\ge 3$ is $\smallO(\delta^2(\varepsilon))$. 

Substituting in the Taylor expansions to (\ref{equation:RES cut-off varepsilon}) for $n=2$ gives
\[\left(1 - \frac{(N - 1) \delta}{\bar S} + \frac{(N - 1)(N-2) \delta^2}{2 \bar S^2}\right)(\varepsilon - \delta) = c \left(\frac{(N-1)\delta}{\bar S} - \frac{2(N-1)(N-2)\delta^2}{3 \bar S^2} \right) + \smallO(\delta^2).
\]
For $n\ge 3$, the cutoff type will be selected to enter the auction with probability $1$ instead of $2/3$ when two other bidders opt in. This is reflected by the coefficient on $p_2$ taking the value of $1$ instead of $2/3$. In this case, substituting the Taylor expansions gives
\[\left(1 - \frac{(N - 1) \tilde\delta}{\bar S} + \frac{(N - 1)(N-2) \tilde\delta^2}{2 \bar S^2}\right)(\varepsilon - \tilde\delta) = c \left(\frac{(N-1)\tilde\delta}{\bar S} - \frac{(N-1)(N-2)\tilde\delta^2}{2 \bar S^2} \right) + \smallO(\tilde\delta^2),
\]
where we use the $\tilde\delta$ to denote the cutoff when $n\ge 3$. If only the terms of order $\delta$ are considered, then there is no difference in the conditions. Therefore the difference of the cutoff signal from changing $n=2$ to $n\ge 3$ is of order $\delta^2$ and we denote $\delta - \tilde\delta = \mathcal{O}(\delta^2)$. We use this fact to compare the relative size of the changes in the \textit{participation effect} and \textit{selection effect} from increasing $n=2$ to $n\ge 3$.

From (\ref{equation:PE}) and substituting in $\delta(\varepsilon) = \bar S - \alpha_0(\varepsilon)$, the change in the \textit{participation effect} is
\begin{align*}
    PE((n,1),\mathcal{X}) - PE((2,1),\mathcal{X}) = g(\delta) - g(\tilde\delta),
\end{align*}
where $g(\delta) = \sum_{k=2}^N q_k(\delta) \left(\bar S - \frac{2\delta}{k+1} + \mathbb{E}[T]\right)$ and $q_k(\delta) = \binom{N}{k}\left(\frac{\delta}{\bar S}\right)^k \left(1 - \frac{\delta}{\bar{S}}\right)^{N-k}$. It follows that $g(\delta) = (\bar S + \mathbb{E}[T]) \frac{N(N-1)}{2}\frac{\delta^2}{\bar S^2} + \mathcal{O}(\delta^3)$ and the difference in participation effect is
\[
g(\delta) - g(\tilde\delta) = \frac{N(N-1)}{2} \frac{\bar S +\mathbb{E}[T]}{\bar S^2} \left(\delta^2 - \tilde\delta^2\right) + \mathcal{O}(\delta^3,\tilde\delta^3).
\]
Since $\delta^2 - \tilde\delta^2 = (\delta - \tilde\delta)(\delta + \tilde\delta)$ its order is $\mathcal{O}(\delta^2) \cdot \mathcal{O}(\delta) = \mathcal{O}(\delta^3)$. Then the order of the change in \textit{participation effect} is also $\mathcal{O}(\delta^3)$.

From Proposition~\ref{proposition:entry restriction} we know that the \textit{selection effect} decreases as $n$ increases for a given environment. So the change in \textit{selection effect} can be bounded by the \textit{selection effect} of $n=2$:

\[
|SE((n,1),\mathcal{X}) - SE((2,1),\mathcal{X})|<SE((2,1),\mathcal{X}) = \sum_{k=3}^N q_k(\delta) \left(\frac{k - 1}{k + 1} - \frac{1}{3}\right) \delta. 
\]
The probability that 3 potential bidders opt in is 
\[
q_3 = \binom{N}{3}\left(\frac{\delta}{\bar S}\right)^3 \left(1 - \frac{\delta}{\bar{S}}\right)^{N-3} = \frac{N(N-1)(N-2)}{6}\left(\frac{\delta}{\bar{S}}\right)^3 + \mathcal{O}(\delta^4),
\]
and all $q_k$ for $k\ge4$ are $\smallO(\delta^3)$. Then the \textit{selection effect} for $n=2$ can be written as
\[
\sum_{k=n+1}^N q_k \left(\frac{k - 1}{k + 1} - \frac{n - 1}{n + 1}\right) \delta = \frac{N(N-1)(N-2)}{36}\left(\frac{\delta^4}{\bar S^3}\right) + \mathcal{O}(\delta^5) = \mathcal{O}(\delta^4),
\]
Therefore the change in \textit{selection effect} from $n=2$ to any $n>2$ is bounded above by a term of order $\delta^4$. This implies that the decrease in the \textit{participation effect} is of higher order than the increase in \textit{selection effect} from restricting entry from $n>2$ to $n=2$. Therefore, for sufficiently high cost of entry that is less than $\bar S + \mathbb{E}[T]$, restricting entry to $n=2$ will increase the expected revenue of the auction.

\subsection*{Proof of Proposition~\ref{proposition:indicative bidding}}
Let $\alpha_1\in[0,\bar S]$ be given. For all $\alpha_0 < \alpha_1$, $p_0 = p_{0,0}$ and $p_t = \sum_{j+k = t} p_{k,j}$ for $t=1,\ldots, N-1$. In the following, we show that $\min\left\{1,\frac{n-j}{k+1}\right\} \le \min\left\{1, \frac{n}{t+1}\right\}$ for $t = j+k$ and that the inequality is strict when $n<j+k+1$.

When $n \ge j+k+1$, then $\min\left\{1,\frac{n-j}{k+1}\right\} = 1$ and $\min\left\{1,\frac{n}{k+j+1}\right\}=1$. Then for $k+j=t$ and $n\ge t+1$, $\min\left\{1,\frac{n-j}{k+1}\right\} = \min\left\{1, \frac{n}{t+1}\right\}$.

When $n \le j+k+1$ and $j+k=t$, then $\min\left\{1,\frac{n-j}{k+1}\right\} = \frac{n-j}{k+1}$ and $\min\left\{1, \frac{n}{t+1}\right\}=\frac{n}{k+j+1}$. Then the difference in these fractions is
\begin{align*}
    \frac{n-j}{k+1} - \frac{n}{k+j+1} = \frac{(n-j)(k+j+1) - n(k+1)}{(k+j+1)(k+1)} = \frac{nj - j(k+j+1)}{(k+j+1)(k+1)} \le 0.
\end{align*}
The last inequality follows from $n \le j+k+1$ and is strict when $n < j+k+1$. It follows that for any $\alpha_1\in[0,\bar S]$
\begin{align}
c \sum_{k=1}^{N-1} p_k \min\left\{1,\frac{n}{k+1}\right\} & \ge c \sum_{j=0}^{n-1} \sum_{k=1}^{N-1-j} p_{k,j} \min\left\{1,\frac{n-j}{k+1}\right\}.\label{ineq:PE}
\end{align}

It follows that the second term of (\ref{RES cut-off}) is less negative than the second term of (\ref{IND cut-off 1}) for any fixed $\alpha_0$. Moreover, each of these terms becomes less negative as $\alpha_0$ increases (see argument in the proof of Proposition~\ref{proposition:PE vs SE}). Because the first term of each equation is the same for any $\alpha_0$ and $\alpha_1<\bar S$, and the terms increase in $\alpha_0$, the value of $\alpha_0$ that satisfies (\ref{RES cut-off}) is higher than the $\alpha_0$ that satisfies (\ref{IND cut-off 1}). Whenever $\alpha_1 < \bar S$ it follows that $p_{k,j}>0$ for some $j,k$ where $n<j+k+1$ and therefore the inequality of (\ref{ineq:PE}) is strict. This implies that $\alpha_0$ is strictly lower when $\overline M = 2$ than when $\overline M = 1$. The \textit{participation effect} only depends on the two-stage mechanism through its equilibrium value of $\alpha_0$. Moreover, it is decreasing in $\alpha_0$ (see (\ref{equation:PE})) and therefore, it is strictly less when $\overline M = 2$ than when $\overline M = 1$.

\newpage
\section{Instructions}\label{appendix:instructions}
    Welcome to our experiment! You will receive 30 RMB for showing up on time. Please read these instructions carefully and completely. Properly understanding the instructions will help you make better decisions and hence earn more money. The experiment will last about 1.5 hours. Your payoff in this experiment will be measured in experimental currency (i.e., EC).  At the end of the experiment, we will convert your payoff in EC to cash and pay you in private. The exchange rate is 1 EC = 1.5 RMB.  

\vspace{5mm}
Your \underline{total payment} from this experiment will be the sum of:

 (1) Your show-up fee: 30 RMB;
 
 (2) Your payoff in this experiment;
\vspace{5mm}

Please keep in mind that you are not allowed to communicate with other participants during the experiment. If you do not obey this rule you will be asked to leave the laboratory. Whenever you have a question, please raise your hand; an experimenter will come to assist you. 

\section*{Your task}
At the beginning of the experiment, you will be given 30 EC as your initial endowment. In this experiment, there are 20 rounds in total. Before round 1, you will be randomly assigned to a group of 5 players and will stay in this group for the first 10 rounds. In round 11, you will be randomly reassigned to a new group of 5 players and stay in that group for the rest of the experiment. The computer will randomly assign letters from A to E to each of you as your player label and this assignment changes in every round. You will only be competing against the participants in your group. 

In each round, you will participate in an auction. The auction takes place in two stages: in stage 1, you decide if you would like to enter stage 2 or not; in stage 2, (if you enter successfully,) you choose how much to bid in the auction. Your total valuation towards the asset that you are bidding for is composed of two parts: You will get the first part of your valuation in stage 1 before you make the entry decision; then you will get the second part of your valuation in stage 2 before you enter your bid if you enter the auction. \vspace{5mm}

\underline{Stage 1: Entry decision}\\
The computer first randomly draws an integer from 1 to 100 (including 1 and 100) for each participant independently. Each integer has 1\% chance to be drawn. You only observe your own draw.  This number reveals the first part of your valuation towards the asset that you are bidding for in stage 2.  After seeing this information, you need to make a decision on whether or not to enter the auction stage based on this partial information you have about your total valuation for the asset. You can choose 1 of the 3 options to indicate your willingness to enter: “0” (Do Not Enter), “1” and “2” (both represent enter with 2 giving you a higher entry priority than 1). 

The number of entrants in each group in the auction stage is restricted to at most 2. If the number of players that choose either “1” or “2” is equal or less than 2, then all participants who choose “1” or “2” are selected to enter automatically. However, if there are more than 2 participants in your group choosing to enter, a selection process will apply. First we randomly select up to two players from those who chose “2”. Then, if less than 2 players have been selected, we randomly select the remaining entrants from those who chose “1”.

The selected entrant(s) will proceed to stage 2 and an entry fee will be charged. \textbf{Notice that the entry fee is 5 EC in Round 1-10, and then increases to 25 EC in Round 11-20.}  If you are not selected, no entry fee will apply, and you do not need to make any more decisions in this round. \vspace{5mm}

\underline{Stage 2: Auction stage}\\
In this stage, all the entrants first see two cards on the screen. Each of the cards has two possible values, 0 and 100. The computer will randomly assign one of the values (0, or 100 with equal chance) independently to each card. The sum of the two numbers assigned to the two cards is the second part of your the asset valuation, which is the same for all entrants. There are four possible scenarios: both cards have 0; the first card is 0 and the second card is 100; the first card has 100 and the second card has 0; both cards have 100. Hence, the second part of the value shared by all entrants could be 0 EC with 25\% chance, 100 EC with 50\% chance, and 200 EC with 25\% chance. 

Each entrant’s full valuation for the asset (V) is calculated as the sum of the first part of the valuation revealed in stage 1 (i.e., the independent private draw between 1-100) and the second part of the valuation (i.e., the sum of the numbers on the two cards which are drawn for each group) shared by all entrants revealed in stage 2.  In the example below, if an entrant’s draw in the first stage is 56, and then the computer assigned 100 and 0 to two cards in stage 2 respectively, his/her total valuation for the asset is 56+(0+100)=156.

\begin{figure}[h!]
    \centering
    \includegraphics[width=0.8\textwidth]{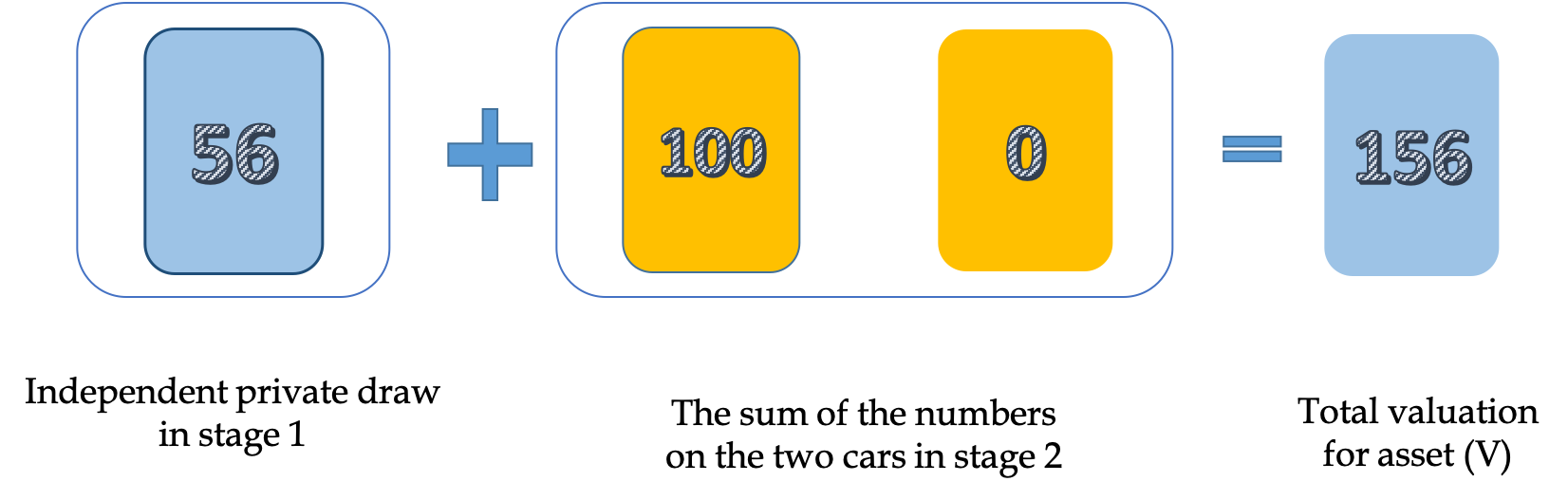}
    \label{fig:screen_shot_total_v}
\end{figure}

After learning your total valuation, all the entrants need to bid for the asset. The entrant with the highest bid in your group will win the asset. The winner will pay an amount equal to the second highest bid. If there is a tie for the highest bid, a winner will be randomly selected among these bidders. In this case, the second highest bid is the same as the highest bid. When there is only one entrant, that player wins the asset and pays zero (i.e. the second highest bid). 

\section*{Your payoff}
Your payoff for each round will be calculated at the end of auction as the following:
\begin{itemize}
    \item If you do not enter stage 2: your payoff = 0
    \item If you choose to enter and are selected:
        \begin{itemize}
            \item If you lose, your payoff= - entry fee
            \item If you win, your payoff= your total valuation - the second highest bid - entry fee
        \end{itemize}
\end{itemize}

\section*{Summary}
You will play 20 rounds of this 2-stage auction game. You will be randomly assigned to a group of 5 participants twice during this experiment (first before round 1 and again before round 11). In each round, you decide whether to enter the auction or not and how much to bid. Your valuation for the asset in each round is determined by the sum of your independent draw (revealed in stage 1) and the common draw given to all entrants in stage 2.  At the end of each round, all group members’ entry decisions, two parts of the valuation information, entrants’ bids and payoffs will be displayed to you, regardless of whether you chose to enter stage 2 or not. See a screenshot of this page below: 

\begin{figure}[h!]
    \centering
    \includegraphics[width=0.8\textwidth]{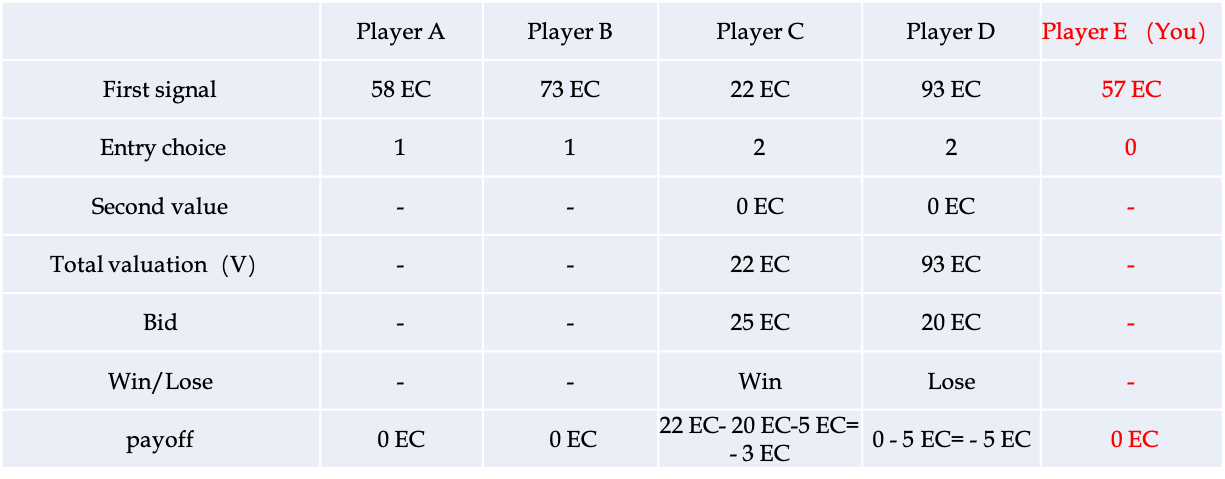}
    \label{fig:screen_shot_summary}
\end{figure}

In addition, a history table which gives information about entry decisions and winner’s payoffs from previous rounds is provided.

\section*{Your total payment}
After you complete all 20 rounds, the computer will randomly draw 1 round to pay you. Your total payoff from the experiment will be your endowment (30 EC) plus your payoff in the drawn round. Your total payoff will then be converted into cash and paid to you together with your show-up fee (30 RMB) at the end of today’s session. 

\section*{Other information}
To further ensure that everyone in this lab understands the game properly, you will need to answer several control questions that are constructed based on the information given out in these instructions. The experiment will start once all of you have answered these questions correctly. Please do not hesitate to ask for help if you have any questions regarding the information provided in our instructions or the control questions we ask you to answer. 

At the end of today’s experiment, you will also need to fill out a small post-experimental questionnaire, including some demographic information (e.g., your gender, age, major...) and your decisions in the experiment. All the information you provide will be kept anonymous and is strictly confidential. The only purpose of collecting this information from you is for academic research analysis.

Thank again for your participation and patience! The experiment will start soon…

\newpage
\section{Additional results}\label{appendix:results}

\renewcommand{\thetable}{C-\arabic{table}}
\setcounter{table}{0}
\renewcommand{\thefigure}{C-\arabic{figure}}
\setcounter{figure}{0}

\begin{threeparttable}
    \footnotesize
\setstretch{1}
  \centering
  \caption{Predicted revenue, \textit{participation} and \textit{selection effects} given drawn signals}
    \begin{tabular}{lccccccc}
    \toprule
          &       & \multicolumn{1}{l}{\textbf{Low-cost (c=5)}} &       &       &       & \multicolumn{1}{l}{\textbf{High-cost (c=25)}} &  \\
    \midrule
          & \textbf{Unr} & \textbf{Res} & \textbf{Ind} &       & \textbf{Unr} & \textbf{Res} & \textbf{Ind} \\
\cmidrule{2-4}\cmidrule{6-8}    
$\Pi_{Pred}(\mathcal{M},\mathcal{X})$ & 153.93 & 145.69 & 150.02 &       & 101.80 & 104.83 & 106.84 \\
   $\hat\Pi_{Pred}(\alpha_0(\mathcal{M}),\mathcal{X})$ & 153.93 & 155.97 & 154.85 &       & 101.80 & 106.86 & 109.00 \\
    $\Pi_{Pred}(\mathcal{X}_0)$ & 168.5 & 166.79 & 164.34 &       & 163.58 & 154.32 & 162.88 \\
    \textit{Participation effect} & 14.57 & 10.82 & 9.50 &       & 61.78 & 47.45 & 53.89 \\
    \textit{Selection effect}   & 0     & 10.28 & 4.83 &       & 0     & 2.04 & 2.15 \\
    \bottomrule
    \end{tabular}
  \label{tab: Pred_given_signal}%
  \begin{tablenotes}[flushleft]
\item Note: $\Pi_{Pred}(\mathcal{M},\mathcal{X})$ is the expected revenue given the signal drawn in the experiment. $\hat\Pi_{Pred}(\alpha_0(\mathcal{M}),\mathcal{X})$ is the expected revenue in the counterfactual scenario where all potential bidders who send a positive entry message enter the auction stage, given the signal drawn in the experiment. $\Pi_{Pred}(\mathcal{X}_0)$ is the auction revenue in the zero-entry-cost environment. `\textit{Participation effect}' is the difference between $\Pi_{Pred}(\mathcal{X}_0)$ and $\hat\Pi_{Pred}(\alpha_0(\mathcal{M}),\mathcal{X})$, and `\textit{Selection effect}' is the difference between $\hat\Pi_{Pred}(\alpha_0(\mathcal{M}),\mathcal{X})$ and $\Pi_{Pred}(\mathcal{M},\mathcal{X})$.
\end{tablenotes}

\end{threeparttable}

\vspace{1cm}

\begin{center}

\begin{threeparttable}
    \footnotesize
\setstretch{1}
    \centering
  \caption{Recalibrated ``observed" revenue, \textit{participation} and \textit{selection} effects}
    \begin{tabular}{lccccccc}
    \toprule
          &       & \multicolumn{1}{l}{\textbf{Low-cost (c=5)}} &       &       &       & \multicolumn{1}{l}{\textbf{High-cost (c=25)}}&  \\
\midrule     & \textbf{Unr} & \textbf{Res} & \textbf{Ind} &       & \textbf{Unr} & \textbf{Res} & \textbf{Ind} \\
\cmidrule{2-4}\cmidrule{6-8}    
$\Pi_{Obs}(\mathcal{M},\mathcal{X})$ & 156.25 & 140.75 & 149.04 &       & 107.10 & 104.98 & 131.89 \\
$\Pi_{EB}(\mathcal{M},\mathcal{X})$ & 149.96 & 135.68& 141.37 &   &101.37 &98.68 &124.86 \\
$\hat\Pi_{Obs}(\alpha_0(\mathcal{M}),\mathcal{X})$  & 149.96 & 151.87 & 153.69 &       & 101.37 & 105.68 & 129.81 \\
$\Pi_{Pred}(\mathcal{X}_0)$  & 168.5 & 166.79 & 164.34 &       & 163.58 & 154.32 & 162.88 \\
    \textit{Participation effect}& 18.54 & 14.93 & 10.65 &       & 62.20 & 48.65 & 33.07 \\
    \textit{Selection effect}   & 0.00 & 16.18 & 12.68  &       & 0.00  & 7.00 & 7.55 \\
    \bottomrule
    \end{tabular}%
  \label{tab:Observed_adjust_bidding}%
\begin{tablenotes}[flushleft]
\item Note: $\Pi_{Obs}(\mathcal{M},\mathcal{X})$ is the observed revenue in the experiment. $\Pi_{EB}(\mathcal{M},\mathcal{X})$ is the recalibrated ``observed" revenue expected in the experiment, which is calculated using observed entry decisions and equilibrium bids, and further accounts for revenue variations caused by the random tie-breaking in \textit{Res} and \textit{Ind} by calculating the expected revenue when demand is higher than the entry cap. $\hat\Pi_{Obs}(\alpha_0(\mathcal{M}),\mathcal{X})$ is the counterfactual scenario where all potential bidders who send a positive entry message enter the auction stage. $\Pi_{Pred}(\mathcal{X}_0)$ is the auction revenue in the zero-entry-cost environment. '\textit{Participation effect}' is  $\Pi_{Pred}(\mathcal{X}_0) - \hat\Pi_{Obs}(\alpha_0(\mathcal{M}),\mathcal{X})$, and `\textit{Selection effect}' is $\hat\Pi_{Obs}(\alpha_0(\mathcal{M}),\mathcal{X}) - \Pi_{EB}(\mathcal{M},\mathcal{X})$. 
\end{tablenotes}
\end{threeparttable}
\end{center}

\vspace{1cm}

\begin{center}

\begin{threeparttable}[htbp]
\footnotesize
\setstretch{1}
  \caption{Multilevel linear regressions on expected revenue given observed entry choice and equilibrium bids}
    \begin{tabular}{lcccc}
    \toprule
          & \multicolumn{2}{c}{Model 1} & \multicolumn{2}{c}{Model 2} \\
\cmidrule{2-5}          & Low-cost & High-cost & Low-cost & High-cost \\
    \midrule
    Unr (Cons.) & 150.0*** & 101.4*** & -13.05*** & -31.02*** \\
          & (5.405) & (6.999) & (4.574) & (8.582) \\
    Res   & -14.28* & -2.687 & -12.62*** & 4.806 \\
          & (7.643) & (9.898) & (3.925) & (8.803) \\
    Ind   & -8.592 & 23.49** & -4.570 & 24.05*** \\
          & (7.643) & (9.898) & (3.926) & (8.797) \\
    V$_{2}$    &       &       & 0.967*** & 0.809*** \\
          &       &       & (0.0216) & (0.0361) \\
    Observations & 720   & 720   & 720   & 720 \\
    Number of groups & 72    & 72    & 72    & 72 \\
    \bottomrule
    \end{tabular}%
  \label{tab:rev_reg_hypothetical}
  \begin{tablenotes}[flushleft]
\item Note: Expected revenue is calculated using equilibrium bids given actual entry decisions, and further accounts for revenue variations caused by the random selection in \textit{Res} and \textit{Ind} when demand is higher than the entry cap. It is the same expected revenue used to calculate the \textit{participation} and \textit{selection} effects. Model 1 and Model 2 are both multilevel linear regressions with random intercepts for sessions and groups, while Model 2 further controls for the second-highest value in the group (V$_{2}$). Standard errors in parentheses are adjusted for clustering at both the session and group levels. *** Significant at the 1\% level, ** at the 5\% level, and * at the 10\% level.
\end{tablenotes}
\end{threeparttable}
\end{center}

\begin{threeparttable}[htbp]
\footnotesize
\setstretch{1}
  \centering
  \caption{Entry choice breakdown}
    \begin{tabular}{lccccccccccccc}
    \toprule
&\multicolumn{6}{c}{\textbf{Low-cost (c=5)}} && \multicolumn{6}{c}{\textbf{High-cost (c=25)}} \\
\midrule
& \multicolumn{2}{c}{\textbf{Unr}}& \multicolumn{2}{c}{\textbf{Res}} & \multicolumn{2}{c}{\textbf{Ind}} && \multicolumn{2}{c}{\textbf{Unr}} & \multicolumn{2}{c}{\textbf{Res}}&\multicolumn{2}{c}{\textbf{Ind}} \\
\cmidrule{2-7}\cmidrule{9-14}& No.   & Pct. & No.& Pct.&No.& Pct. && No.   & Pct. & No.   & Pct. & No.   & Pct. \\
\textbf{Nash Entry} & \textbf{931} & \textbf{78\%} & \textbf{917} & \textbf{76\%} & \textbf{625} & \textbf{52\%} && \textbf{990} & \textbf{83\%} & \textbf{958} & \textbf{80\%} & \textbf{632} & \textbf{53\%} \\
(0,0) & 405& 34\% & 313 & 26\% & 260 & 22\% && 668   & 56\% & 597   & 50\% & 514   & 43\% \\
(1,1)& 526   & 44\% & 604   & 50\% &134& 11\% && 322 & 27\% & 361   & 30\% & 118   & 10\% \\
(2,2)&&&&& 231& 19\% &&&&&&&  \\
\textbf{Under Entry} & \textbf{132} & \textbf{11\%} & \textbf{141} & \textbf{12\%} & \textbf{105} & \textbf{9\%} && \textbf{99} & \textbf{8\%} & \textbf{95} & \textbf{8\%} & \textbf{86} & \textbf{7\%} \\
(0,1) & 132& 11\% & 141& 12\% & 97 & 8\% && 99& 8\% & 95& 8\% & 86& 7\% \\
(0,2) &&&&& 8 & 1\% &&&&&&&  \\
\textbf{Over Entry} & \textbf{137} & \textbf{11\%} & \textbf{142} & \textbf{12\%} & \textbf{169} & \textbf{14\%} && \textbf{111} & \textbf{9\%} & \textbf{147} & \textbf{12\%} & \textbf{209} & \textbf{17\%} \\
(1,0)& 137& 11\% & 142& 12\% & 89 & 7\% && 111& 9\% & 147   & 12\% & 116   & 10\% \\
(2,0) &&&&& 80& 7\% &&&&&& 93    & 7\% \\
\textbf{Message Misuse}  &  &  & &  & \textbf{301} & \textbf{25\%} &&  & &  &  & \textbf{273} & \textbf{23\%} \\
(1,2)&&&&& 6& 0.5\% &&&&&&&  \\
(2,1) &&&&& 295& 24.5\% &&&&       &       & 273   & 23\% \\
 N & 1200  && 1200  && 1200  &&& 1200  & & 1200  & & 1200  &  \\

    \bottomrule
    \end{tabular}%
  \label{tab:entry_breakdown}
 \begin{tablenotes}[flushleft]
\item Note: We classify participants' entry choices into four categories: Nash entry, over-entry, under-entry, and (positive entry) message misuse. Nash entry, over-entry, and under-entry refer to cases in which the chosen message matches, exceeds, or falls below the equilibrium message implied by the participant’s initial signal, respectively. Message misuse captures cases in which the chosen positive message does not match the equilibrium positive message. Within each category, we further disaggregate outcomes by the ordered pair (chosen message, equilibrium message); the first element is the chosen message and the second is the equilibrium message.
\end{tablenotes}
\end{threeparttable}

\begin{figure}[htbp]
    \centering
    \includegraphics[width=0.9\textwidth]{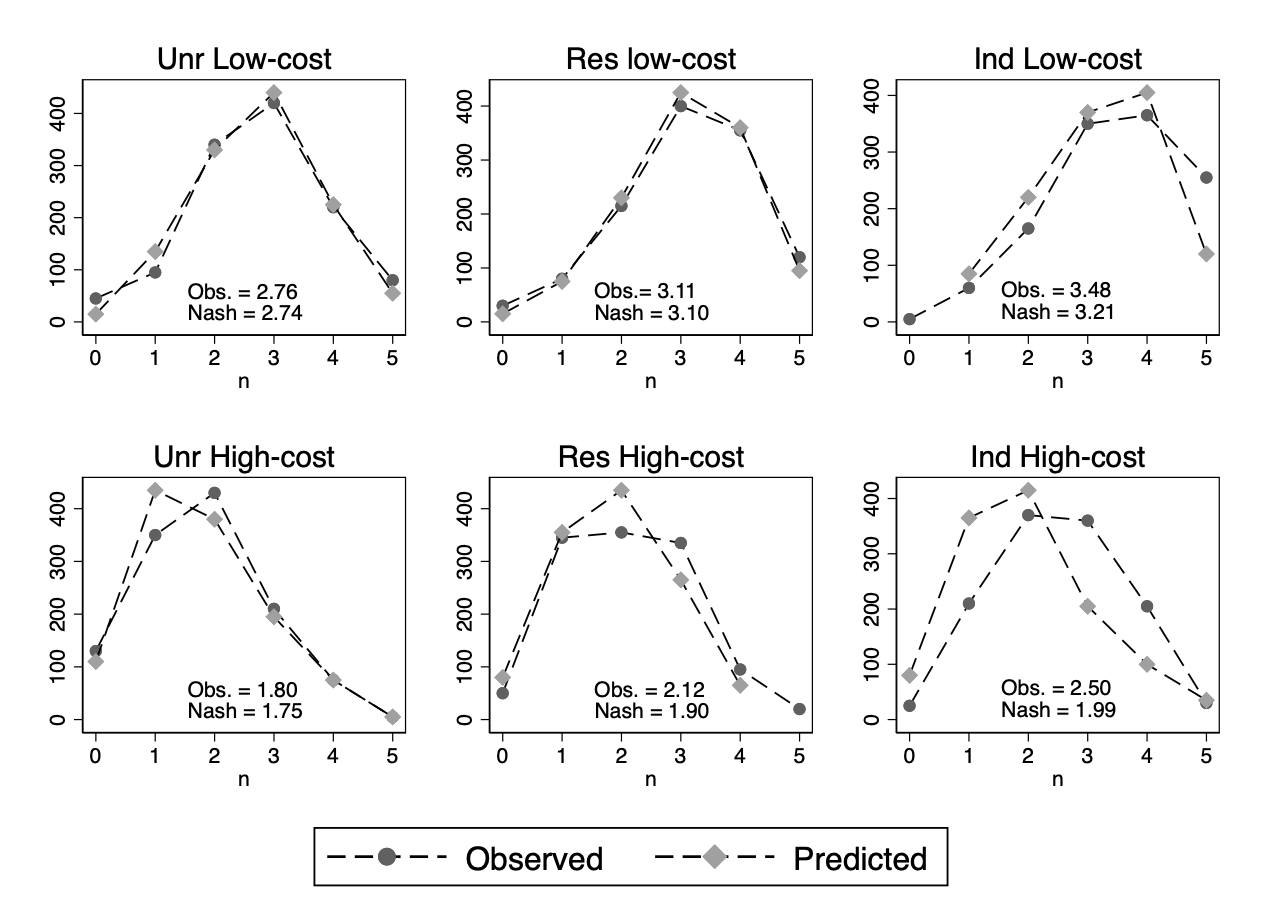}
    \caption{Frequency of the average number of participants who chose to enter}
    \label{fig:freq_n}
\end{figure}

\begin{table}[htbp]
  \centering
  \begin{threeparttable}
  \footnotesize
      \caption{Entry choice: Multilevel ordered logistic regressions}
        \begin{tabular}{lccc}
        \toprule
              & Unr   & Res   & Ind \\
        \midrule
        $S_i$  & 0.0884*** & 0.0920*** & 0.0926*** \\
              & (0.00404) & (0.00440) & (0.00343) \\
        \textit{i.low-cost} & 1.593*** & 1.741*** & 2.071*** \\
              & (0.261) & (0.337) & (0.366) \\
        Cutoff 1 & 5.538*** & 5.163*** & 4.457*** \\
              & (0.299) & (0.333) & (0.310) \\
        Cutoff 2 &       &       & 6.517*** \\
              &       &       & (0.338) \\
        Observations & 2,400 & 2,400 & 2,400 \\
        Number of individuals & 120   & 120   & 120 \\
        \bottomrule
        \end{tabular}%
      \label{tab:entrychoice_melogit}%
        \begin{tablenotes}[flushleft]
        \item Note: We estimate entry choice using a multilevel ordered logistic model with random intercepts at the group and individual levels for each mechanism. $S_i$ is the initial signal, \textit{i.low-cost} takes the value of 1 in the \text{low-cost} condition, and 0 in the \textit{high-cost} condition. Cutoff 1 and Cutoff 2 are the estimated cutoffs from the multilevel ordered logistic model. Standard errors in parentheses are adjusted for clustering at both the group and individual levels. *** Significant at the 1\% level, ** at the 5\% level, and * at the 10\% level.
        \end{tablenotes}
  \end{threeparttable}
\end{table}%

\begin{table}[htbp]
\begin{center}
    \begin{threeparttable}
  \footnotesize
  \caption{Bidders' profit: summary statistics}
    \begin{tabular}{rcccrccc}
    \toprule
          & \multicolumn{3}{c}{\textbf{Low-cost}} &       & \multicolumn{3}{c}{\textbf{High-cost}} \\
\cmidrule{1-4}\cmidrule{6-8}    \multicolumn{1}{l}{Unr} & \textbf{Obs} & \textbf{Predicted} & \textbf{NE} &       & \textbf{Obs} & \textbf{Predicted} & \textbf{NE} \\
          & 0.96  & 14.50 & 15.71 &       & 7.03  & 24.19 & 12.07 \\
          & \textcolor[rgb]{ .682,  .667,  .667}{(60.20)} & \textcolor[rgb]{ .682,  .667,  .667}{(52.09)} & \textcolor[rgb]{ .682,  .667,  .667}{} & \textcolor[rgb]{ .682,  .667,  .667}{} & \textcolor[rgb]{ .682,  .667,  .667}{(102.98)} & \textcolor[rgb]{ .682,  .667,  .667}{(102.96)} & \textcolor[rgb]{ .682,  .667,  .667}{} \\
    \multicolumn{1}{l}{Res} & \textbf{Obs} & \textbf{Predicted} & \textbf{NE} &       & \textbf{Obs} & \textbf{Predicted} & \textbf{NE} \\
          & 16.77 & 20.92 & 20.95 &       & 11.3  & 17.57 & 14.27 \\
          & \textcolor[rgb]{ .682,  .667,  .667}{(58.86)} & \textcolor[rgb]{ .682,  .667,  .667}{(45.75)} & \textcolor[rgb]{ .682,  .667,  .667}{} & \textcolor[rgb]{ .682,  .667,  .667}{} & \textcolor[rgb]{ .682,  .667,  .667}{(90.92)} & \textcolor[rgb]{ .682,  .667,  .667}{(86.01)} & \textcolor[rgb]{ .682,  .667,  .667}{} \\
    \multicolumn{1}{l}{Ind} & \textbf{Obs} & \textbf{Predicted} & \textbf{NE} &       & \textbf{Obs} & \textbf{Predicted} & \textbf{NE} \\
          & 11.26 & 21.32 & 17.16 &       & -8.29 & 21.88 & 14.27 \\
          & \textcolor[rgb]{ .682,  .667,  .667}{(55.61)} & \textcolor[rgb]{ .682,  .667,  .667}{(48.02)} & \textcolor[rgb]{ .682,  .667,  .667}{} & \textcolor[rgb]{ .682,  .667,  .667}{} & \textcolor[rgb]{ .682,  .667,  .667}{(82.83)} & \textcolor[rgb]{ .682,  .667,  .667}{(93.16)} & \textcolor[rgb]{ .682,  .667,  .667}{} \\
    \bottomrule
    \end{tabular}%
  \label{tab:bid_profit_sum}%
\begin{tablenotes}[flushleft]
\item Note: Predicted bidders' profit is the expected bidders' profit based on the drawn values in the experiment. NE is the theoretical expected bidders' profit (not based on drawn values). Standard deviations are in parentheses. 
\end{tablenotes}
   \end{threeparttable}
 \end{center}
\end{table}

\begin{table}[htbp]
\begin{center}
    \begin{threeparttable}
  \footnotesize
 \caption{Social welfare: summary statistics}
    \begin{tabular}{rcccrccc}
    \toprule
          & \multicolumn{3}{c}{\textbf{Low-cost}} &       & \multicolumn{3}{c}{\textbf{High-cost}} \\
\cmidrule{1-4}\cmidrule{6-8}    \multicolumn{1}{l}{Unr} & \textbf{Obs} & \textbf{Predicted} & \textbf{NE} &       & \textbf{Obs} & \textbf{Predicted} & \textbf{NE} \\
\cmidrule{2-4}\cmidrule{6-8}          & 157.22 & 168.43 & 167.10 &       & 114.12 & 125.99 & 121.96 \\
          & (77.11) & \textcolor[rgb]{ .682,  .667,  .667}{(71.73)} & \textcolor[rgb]{ .682,  .667,  .667}{} & \textcolor[rgb]{ .682,  .667,  .667}{} & (82.32) & \textcolor[rgb]{ .682,  .667,  .667}{(82.54) } & \textcolor[rgb]{ .682,  .667,  .667}{} \\
    \multicolumn{1}{l}{Res} & \textbf{Obs} & \textbf{Predicted} & \textbf{NE} &       & \textbf{Obs} & \textbf{Predicted} & \textbf{NE} \\
          & 157.52 & 166.6 & 167.74 &       & 116.28 &  122.40  & 131.09 \\
          & (81.46) & \textcolor[rgb]{ .682,  .667,  .667}{(78.14)} & \textcolor[rgb]{ .682,  .667,  .667}{} & \textcolor[rgb]{ .682,  .667,  .667}{} & (76.52) & \textcolor[rgb]{ .682,  .667,  .667}{(76.74)} & \textcolor[rgb]{ .682,  .667,  .667}{} \\
    \multicolumn{1}{l}{Ind} & \textbf{Obs} & \textbf{Predicted} & \textbf{NE} &       & \textbf{Obs} & \textbf{Predicted} & \textbf{NE} \\
          & 160.3 & 171.34 & 172.02 &       & 123.6 &  128.72 & 131.09 \\
          & (74.20) & \textcolor[rgb]{ .682,  .667,  .667}{(71.22)} & \textcolor[rgb]{ .682,  .667,  .667}{} & \textcolor[rgb]{ .682,  .667,  .667}{} & (75.65) & \textcolor[rgb]{ .682,  .667,  .667}{(79.26)} & \textcolor[rgb]{ .682,  .667,  .667}{} \\
    \bottomrule
    \end{tabular}%
  \label{tab:social_walfare_sum}%
\begin{tablenotes}[flushleft]
\item Note: Predicted social welfare is the expected social welfare based on the drawn values in the experiment. NE is the theoretical expected social welfare (not based on drawn values). Standard deviations are in parentheses. 
\end{tablenotes}
   \end{threeparttable}
 \end{center}
\end{table}

\clearpage


\end{appendices}

\end{document}